\documentclass[12pt]{iopart}

\pdfoutput=1

\usepackage{fancyhdr}

\usepackage{bm}
\usepackage[english]{babel}
 \expandafter\let\csname equation*\endcsname\relax
 \expandafter\let\csname endequation*\endcsname\relax
\usepackage{amsmath,
amssymb,amsfonts,latexsym}
\usepackage{graphicx}
\usepackage{color}
\usepackage{iopams}
\usepackage{subfigure}
\usepackage{afterpage}
\usepackage{blkarray}
\usepackage{cite}
\usepackage{tikz}
\usetikzlibrary{snakes}

\usepackage{times}
\usepackage[colorlinks,citecolor=blue,linkcolor=red,urlcolor=blue]{hyperref}

\usepackage{cancel}

\newcommand{\be}{\begin{equation}}
\newcommand{\ee}{\end{equation}}
\newcommand{\bea}{\begin{eqnarray}}
\newcommand{\eea}{\end{eqnarray}}

\usepackage{etoolbox}

\makeatletter
\def\@mkboth#1#2{}
\newlength\appendixwidth
\preto\appendix{\addtocontents{toc}{\protect\patchl@section}}
\newcommand{\patchl@section}{%
  \settowidth{\appendixwidth}{\textbf{Appendix }}%
  \addtolength{\appendixwidth}{1.5em}%
  \patchcmd{\l@section}{1.5em}{\appendixwidth}{}{\ddt}%
}

\begin{document}

\title{Entanglement entropy of two disjoint intervals and the recursion formula for conformal blocks } 
\author{Paola Ruggiero$^1$, Erik Tonni$^1$, Pasquale Calabrese$^{1,2}$}
\address{$^1$ SISSA and INFN, via Bonomea 265, 34136 Trieste, Italy}
\address{$^2$ International Centre for Theoretical Physics (ICTP), I-34151, Trieste, Italy}
\date{\today}

\begin{abstract}
We reconsider the computation of the entanglement entropy of two disjoint intervals in a (1+1) dimensional conformal field theory 
by  conformal block expansion of the 4-point correlation function of twist fields. 
We show that accurate results may be obtained by taking into account several terms in the operator product expansion (OPE) of twist fields 
and by iterating the Zamolodchikov recursion formula for each conformal block.
We perform a detailed analysis for the Ising conformal field theory and for the free compactified boson.
Each term in the conformal block expansion can be easily analytically continued and so this approach 
also provides a good approximation for the von Neumann entropy.
\end{abstract}

\newpage 

\tableofcontents

\section{Introduction}

The study of the entanglement content of extended quantum systems such as quantum field theories and many-body systems in
condensed matter became a central research field which allowed to uncover many new features about the fundamental laws of 
nature, see e.g. Refs. \cite{amico-2008, calabrese-2009, eisert-2010, rev-lafl} as excellent reviews and introduction to the subject.
For a pure quantum state $|\psi \rangle$, the bipartite entanglement is measured 
by the entanglement entropy, which is defined as follows. 
We consider a bipartition of the entire system in a subsystem $A$ and its complement $\bar{A}$ and we
introduce the reduced density matrix of the subsystem $A$ as $\rho_A \equiv {\rm Tr}_{\bar{A}} \rho$, where
$\rho\equiv | \psi \rangle \langle \psi |$ is the density matrix of the entire system. 
Then the entanglement entropy between $A$ and $\bar{A}$ is defined as the von Neumann entropy of $\rho_A$:
\begin{equation}
S_A  \equiv  - {\rm Tr} \rho_A \ln \rho_A . 
\end{equation} 
It turned out that the entanglement entropy  provides fundamental information about the systems, especially when dealing with a
conformal field theory (CFT) \cite{hlw-94, vlrk-03, cc-04,cc-09,DCC,abs-11}.

In a generic quantum field theory, the entanglement entropy may be computed via a replica trick \cite{cc-04}, that 
works as follows: one first calculates the moments of $\rho_A$ (i.e. ${\rm Tr} \rho_A^n$ for integer values of $n$) and then  
gets the entanglement entropy as
\begin{equation}
S_A = -  \lim_{n \to 1} \frac{\partial}{\partial n} {\rm Tr} \rho_A^n  ,
\end{equation}
once $n$ has been analytically continued to arbitrary complex values. 
However, the knowledge of all the moments of the reduced density matrix not only gives the entanglement entropy, but 
provides information about the entire spectrum of $\rho_A$ \cite{cl-08}, which is known as entanglement spectrum \cite{lh-08}.   

In the ground state of a (1+1) dimensional quantum field theory, the moments ${\rm Tr} \rho_A^n$ may be computed in the 
path integral formalism and they are equivalent to partition functions $\mathcal{Z}_n$ of the theory on an $n$-sheeted Riemann 
surface \cite{cc-04,cc-09}. 
Moreover, one can equivalently work in a replicated (or $n$-copy) theory, where, instead of having a single field $\varphi$ living 
on the Riemann surface, one works with $n$ non-interacting copies $(\varphi_1, \cdots , \varphi_n)$ living on the complex 
plane but with appropriate boundary conditions at the entangling surface.
In such a theory,  the \emph{branch point twist fields} $ \mathcal{T}_n$ and $\tilde{\mathcal{T}}_n$ are introduced \cite{cc-04,DCC}.
These local fields implement the structure of the Riemann surface through the monodromy conditions of the fields $\{ \varphi_i \}$.
In this way the problem of computing ${\rm Tr} \rho_A^n$ is mapped to the problem of computing correlation functions of such fields. 
If the subsystem $A$ consists of $N$ intervals, $A= \bigcup_{i=1}^N ( u_i, v_i) $, then the moments can be written as 
$2N$-point functions:
\begin{equation}
{\rm Tr} \rho_A^n =
\Big \langle \prod_{i=1}^N \mathcal{T}_n (u_i, 0) \tilde{\mathcal{T}}_n (v_i,  0)  \Big\rangle.
\label{tf-def}
\end{equation}
Twist fields are particularly useful in the context of (1+1) dimensional CFT, where they transform under conformal 
transformations as primary fields with scaling dimension $\Delta_n \equiv \Delta_{\mathcal{T}_n} = \Delta_{\tilde{\mathcal{T}}_n}  =c/24 (n-1/n)$  \cite{cc-04,cc-09}. 
Therefore in the case of a single interval (i.e. $N=1$) embedded in the infinite line, 
the determination of the moments is equivalent to a 2-point function 
whose form is completely fixed by global conformal invariance to be \cite{cc-04}
\begin{equation}
{\rm Tr} \rho_A^n = 
c_n \ell^{- \frac{c}{6}  \left( n -  \frac{1}{n} \right) },
\end{equation}
where $\ell = |v_i - u_i|$ is the length of the interval, $c$ is the central charge of the CFT, and $c_n$ is a non-universal 
normalisation factor (see however \cite{fcm-11}). 
In the replica limit, this leads to the famous formula $S_A=(c/3) \ln \ell $ \cite{hlw-94,cc-04}.

The calculations are much more complicated in the case of more disjoint intervals. 
Indeed, global conformal invariance does not completely fixes the twist fields correlation functions for $N\geq 2$. 
For example, for $N=2$ the required 4-point correlation function can be written as
\begin{equation}
{\rm Tr} \rho_A^n = c_n^2 \left( \frac{(u_2 - u_1)( v_2 - v_1 )}{(v_1 - u_1 )(v_2 - u_2)(v_2- u_1)(v_1 - u_2)}  \right)^{\frac{c}{6} (n- 1/n) } \mathcal{F}_n (x),
\end{equation}
where ${\cal F}_n(x)$ is a model dependent {\it universal} function of the cross ratio
\begin{equation}
x\equiv\frac{(u_1-v_1)(u_2-v_2)}{(u_1-u_2)(v_1-v_2)}\,.
\label{4pR}
\end{equation}

A large literature has been devoted to the  analytical and numerical determination of the functions $\mathcal{F}_n (x)$ 
\cite{GR,cc-04,ch-05,caraglio-2008, fps-09,headrick, CCT-1, CCT, cmv-11,casini-2009, facchi-2008, ATC-Ising, ATC-CB, igloi-2010, fagotti-2010, calabrese-2010, fagotti-2012,cz-13,ctt-14,aef-14,ctc-15,lz-16,ll-16,bkz-17,mmw-18,dei-18,german-18}, but exact results are known only for few models. 
However, even in those few cases when ${\cal F}_n(x)$ is analytically known for arbitrary integer $n$, 
the analytic continuation of the parameter $n$ from integers to complex values, needed in order to get the entanglement entropy,
is a very hard unsolved problem. 
This is due to the fact that the $n$ dependence of the function $\mathcal{F}_n(x)$ is extremely complicated 
(two examples will be reported in the following). Some results are also known for disconnected regions in higher 
dimensions \cite{cft-high-dims}, also in holographic settings \cite{RT,hol}.

A viable and practical way to to overcome the difficulties in the analytic continuation has been proposed in Refs. \cite{ahjk-14,dct-15} 
and consists in performing the continuation numerically by proper rational interpolations for several values of $n$.
While this technique may provide rather accurate results in some instances \cite{dct-15}, it is definitively not satisfactory from 
a theoretical point of view and we would prefer to have an analytic handle on the analytic continuation. 
To this aim, an alternative is to consider an expansion of the function $\mathcal{F}_n (x)$ in which each term shows 
a manageable  dependence on $n$, with a feasible analytic continuation. 
This has been considered e.g. in \cite{CCT}, where a general expression for the expansion in powers of the parameter $x$ has 
been worked out. 
Unfortunately, this expansion generically converges slowly, so that it is very difficult, if not impossible, to get a reliable 
approximation of the entanglement entropy for all values of $x \in [0, 1]$. 
In \cite{GR} Gliozzi and Rajabpour suggested that the expansion in powers of the elliptic variable $q(x)$ (see below for a 
definition) provides an accurate approximation of the entanglement entropy already at the lowest order. 
The main idea was to use of the fusion algebra of twist fields (already introduced in \cite{CCT}) and consider a 
conformal blocks expansion \cite{BPZ, DiFrancesco}, as usually done to deal with 4-point functions. 
Each conformal block is obtained from the recursion formula originally proposed by Al. B. Zamolodchikov \cite{Zam}, 
which is an expansion in $q(x)$ and provides an extremely rapid convergence for the conformal block itself.
We must mention that other systematic expansions have also been considered, but focusing on the semiclassical limit of 
conformal blocks (i.e. in the limit of large central charge) and its relation to the holographic result \cite{Hartman, Sinha, Bin-Chen}.
Furthermore, the Zamolodchikov recursion formula has been used in Ref. \cite{kt-18} to study the time evolution of the entanglement 
entropy starting from locally excited states for large central charge.

In this work, we reconsider the technique introduced in \cite{GR} and we show how the results obtained there may be improved 
by including more \emph{fusion channels}, i.e., further conformal families in the OPE of twist fields, and, when possible, 
by considering a better approximation of each conformal block through the iteration of the Zamolodchikov recursion formula. 

The paper is organised as follows. In Section \ref{section-CBE}, we start by recalling the main steps needed for the expansion in conformal blocks of a generic 4-point function of a  CFT (Section \ref{subs-CBE1}), and we then generalise to the case of twist fields
(Section \ref{subs-CBE2}). 
In Section \ref{section-Zam} we introduce the Zamolodchikov recursion relation, stressing some of its properties and
discussing the approximations that we are going to use. 
In Section \ref{overview} we summarise some known results about  the 4-point twist-field correlations which we need 
as a reference to test the truncations of conformal  block expansion. 
In Section \ref{section-Ising} and \ref{section-CB} we apply this technique to the computation of ${\rm Tr} \rho_A^n$ for the 
Ising conformal field theory and for the compactified boson respectively.
For these two models, we obtain for generic $n$ analytic expansions of the functions $\mathcal{F}_n(x)$  
which approximate the exact results. 
These expansions have a simple $n$ dependence, so that the analytic continuation $n\to1$ can be straightforwardly worked out.
The resulting predictions for the entanglement entropy reasonably match available numerical results.
In Section \ref{section-Conclusions} we critically discuss our findings and stress some unsolved issues deserving further investigation.

\section{Conformal blocks expansion of twist fields correlation functions}  \label{section-CBE}

\subsection{Standard conformal blocks expansion: main steps for the derivation.} \label{subs-CBE1}

In  a generic CFT, global conformal invariance fully fixes the dependence on the positions of the operators in the 2-point and 3-point functions, but
the 4-point correlation is only fixed up to a function of the cross ratios
\begin{equation}
x \equiv \frac{z_{12} z_{34}}{z_{13} z_{24}} ,
\qquad \bar{x} \equiv \frac{\bar{z}_{12} \bar{z}_{34}}{\bar{z}_{13} \bar{z}_{24}}.
\end{equation}
In fact, making use of a Moebius transformation, which maps four generic points as 
\begin{equation}
(z_1, z_2 , z_3 , z_4)  \; \to \; (\infty, 1,  x, 0),
\end{equation}
the correlation of four generic scaling (quasi-primary) fields 
\begin{equation}
\langle \phi_1 (z_1, \bar{z}_1 )  \phi_2 (z_2, \bar{z}_2)  \phi_3 (z_3, \bar{z}_3)  \phi_4 (z_4, \bar{z}_4)  \rangle,
\end{equation}
can be related to the function 
\begin{eqnarray} \label{afterMoebius}
\tilde{\mathcal{F}} (x, \bar{x})& \equiv & \lim_{w, \bar{w} \to \infty}  w^{2 \Delta_1} \bar{w}^{2 \bar{\Delta}_1}  \langle \phi_1 ( w, \bar{w})  \phi_2 (1, 1)  \phi_3 (x, \bar{x})  \phi_4 (0, 0)  \rangle \nonumber  \\
&=& \langle \Delta_1, \bar{\Delta}_1 |  \phi_2 (1, 1)  \phi_3 (x, \bar{x}) | \Delta_4, \bar{\Delta}_4 \rangle ,
\end{eqnarray}
which 
is not fixed by global conformal invariance, but  depends on the dynamical input specifying 
the theory, i.e. the structure constants $C_{ij}^k$ or equivalently the OPE coefficients of primary fields.
Furthermore, $\tilde{\mathcal{F}} (x, \bar{x}) $ can be written as a sum of \emph{conformal blocks} as  \cite{BPZ, DiFrancesco}
\begin{equation} \label{blocks_expansion}
\tilde{\mathcal{F}} (x, \bar{x}) =  \sum_p C_{12}^p C_{34}^p \tilde{F} (x, c, \Delta_p,   \boldsymbol{\Delta} ) \tilde{F} (\bar{x}, c, \bar{\Delta}_p, \bar{\boldsymbol{\Delta}} ),
\end{equation}
where $\boldsymbol{\Delta} \equiv \{ \Delta_1, \Delta_2, \Delta_3,  \Delta_4 \}$, the sum is over all the primary fields of the theory and $\tilde{F} (x,c, \Delta_p,   \boldsymbol{\Delta} )$ are the conformal blocks.

The crucial ingredient to prove (\ref{blocks_expansion}) is the fact that the fields form an algebra, i.e. for any pair of fields we can 
write an operator product expansion
\begin{equation} \label{OPE}
\phi_3 (x, \bar{x}) \phi_4 (0, 0) = \sum_p \frac{C_{34}^p}{x^{\Delta_p - \Delta_3 - \Delta_4} { \bar{x}^{\bar{\Delta}_p - \bar{\Delta}_3 - \bar{\Delta}_4}}  }  \phi_p (0, 0),
\end{equation}
where $\{  \phi_k \}$ is a basis of scaling fields. 
Moreover, since the scaling fields can be collected in conformal families denoted as $[\phi_p]$ (i.e. the set formed by a 
primary field $\phi_p$ and all its descendants), Eq. \eqref{OPE} can be rewritten as
\begin{equation} 
\phi_3 (x, \bar{x}) \phi_4 (0, 0) = \sum_p \frac{C_{34}^p}{x^{\Delta_p - \Delta_3 - \Delta_4} { \bar{x}^{\bar{\Delta}_p - \bar{\Delta}_3 - \bar{\Delta}_4}}  }  [\phi_p (0, 0)],
\end{equation}
where we used the proportionality between the OPE coefficients of a primary operator with its own descendants.
Plugging this OPE into Eq. (\ref{afterMoebius}), we get
\begin{eqnarray}
\tilde{\mathcal{F}} (x, \bar{x}) &=&  \sum_p \frac{C_{34}^p}{  x^{\Delta_p - \Delta_3 - \Delta_4} { \bar{x}^{\bar{\Delta}_p - \bar{\Delta}_3 - \bar{\Delta}_4}}    }   \langle \Delta_1, \bar{\Delta}_1  |  \phi_2 (1, 1)  [\phi_p (0, 0)] |0 \rangle.
\end{eqnarray}
Exploiting the factorisation in  holomorphic and antiholomorphic parts,
the comparison of the above equation with Eq. \eqref{blocks_expansion} determines the conformal block
\begin{equation}
\tilde{F} (x, c, \Delta_p,  \boldsymbol{\Delta})  \equiv [C^p_{12}]^{-1/2} \frac{ \langle \Delta_1, \bar{\Delta}_1 | \phi_2 (1, 1)  [\phi_k (0, 0)] | 0 \rangle|_{hol}}{x^{\Delta_p - \Delta_3 - \Delta_4}},
\end{equation}
and analogously for the antiholomorphic term.

In Eq. \eqref{blocks_expansion}, 
the sum over $p$ is a sum over conformal families showing that the only independent OPE coefficients are the ones of the primary fields. 
In particular, the contribution of the descendants is encoded in the conformal block $\tilde{F} (x)$. 
If one  knows all the proportionality constants relating the OPE coefficients of primaries and their descendants, 
the conformal block can be computed explicitly, but their computation is not generically feasible. 
As we shall show, it is instead convenient to exploit the property that the conformal blocks only depends on few 
variables ($\Delta_p,  \boldsymbol{\Delta}, c$), which are the true dynamical inputs. 

In the case we are interested in,  the four fields have the same scaling dimension $\Delta_i= \bar{\Delta}_i = \Delta$ and the points 
$z_i$ are real $z_i = \bar{z}_i$ (implying $x= \bar{x}$). Hence, we get the simplified expression for the 4-point correlation function
\begin{eqnarray} \label{4CorrFunction}
\langle \phi_1 (z_1, \bar{z}_1 )  \phi_2 (z_2, \bar{z}_2)  \phi_3 (z_3, \bar{z}_3)  \phi_4 (z_4, \bar{z}_4)  \rangle =  \left| \frac{z_{13} z_{24} }{z_{14} z_{23} z_{12} z_{34}} \right|^{4 \Delta} \mathcal{F} (x) ,
\end{eqnarray}
where, using the freedom we have on the definition of the function of the cross ratio, and according to the convention used for the prefactor in \cite{CCT} (which we are going to use in the following), we defined
\begin{equation} \label{tildeF-F}
\mathcal{F} (x) \equiv [x (1-x)]^{ 4\Delta/3}  \tilde{\mathcal{F}} (x) .
\end{equation}
With this notation, the conformal blocks expansion from \eqref{blocks_expansion} and \eqref{tildeF-F} we get
\begin{equation} \label{Fn_convention}
\mathcal{F} (x) =  \sum_p C_{12}^p C_{34}^p F (x, c,\Delta_p,  \boldsymbol{\Delta}) F (x,c, \bar{\Delta}_p ,  \bar{\boldsymbol{\Delta}} ).
\end{equation}

\subsection{Fusion algebra of twist fields and generalised  conformal block expansion} \label{subs-CBE2}

In this manuscript  we are interested in the entanglement entropy of two disjoint intervals which is a 4-point correlation function 
of twist fields (cf. Eq. \eqref{tf-def}).
Since under conformal transformations twist fields behave like primary fields, we expect that
the conformal block expansion could be applied to such a correlation function.
%
The key ingredient is the operator algebra of the twist fields which have an OPE with a generalised form derived in \cite{CCT}
and which reads 
\begin{equation} \label{OPETT}
\mathcal{T}_n (z) \tilde{\mathcal{T}}_n (w) = \sum_{\{k_j \} } C_{k_j} \bigotimes_{j=1}^n \phi_{k_j} \left( \frac{z+w}{2} \right) ,
\end{equation}
where the sum is over the scaling fields $\{\phi_{k_1} \otimes \cdots \otimes \phi_{k_n} \}$ 
of the $n$-copy Hilbert space $\mathcal{H}^{\rm tot} \equiv \otimes_{j=1}^n \mathcal{H}_j$.
Eq. (\ref{OPETT}) tells that the monodromy of the product $\mathcal{T}_n (z) \tilde{\mathcal{T}}_n (w)$ does not affect the state for distance much larger than $|z-w|$
and therefore it is possible to expand $\mathcal{T}_n (z) \tilde{\mathcal{T}}_n (w)$ in a basis of the fields of the $n$-copy theory 
(where we have $n$ identical decoupled fields).

One could then classify the fields entering the OPE according to the global symmetries of the theory.
In fact, the theory we are dealing with, defined by $n$ copies of the \emph{mother} CFT, of central charge $c$, is itself a CFT with central charge $nc$, being invariant under conformal transformations generated by the total stress-energy tensor $T = \sum_{i=1}^n T^{i}$ (the sum of the stress tensors of each replica);
the associated Virasoro generators are the modes of $T$, i.e., $L_k^{\rm tot} = \sum_{j=1}^{n} L_k^{(j)} $. 
Therefore, in the expansion in conformal blocks of any 4-point correlation function for this CFT, each block will include 
the contribution of a primary operator and its descendants with respect to this total Virasoro algebra. 
Thus, Eq. \eqref{OPETT} can be recast in the form
\begin{equation} \label{OPETTcnzn}
\mathcal{T}_n (z) \tilde{\mathcal{T}}_n (w) = \sum_{\alpha } C_{\alpha} [\Phi_{\alpha}] + \cdots,
\end{equation}
where $\Phi_{\alpha}$ are primary fields with respect to the total Virasoro algebra, i.e., they are defined by the property
\begin{equation} \label{primary_tot}
L^{\rm tot}_ m \Phi_{\alpha} = 0 \quad \forall m > 0,
\end{equation}
and $C_{\alpha}$ are the associated OPE coefficients, that can be computed with a method introduced in \cite{CCT} (i.e.  through the computation of $n$-point correlation function of the primaries on the $n$-sheeted Riemann surface, see \cite{CCT} for details), generalized in \cite{cz-13} to deal with general primaries of the theory.

Moreover, for a generic number of intervals it holds $\mathcal{Z}_n \equiv \langle  \mathcal{T}_n \tilde{\mathcal{T}}_n\cdots  \mathcal{T}_n \tilde{\mathcal{T}}_n \rangle$, thus the correlation functions of twist fields have the same symmetries of the partition function $\mathcal{Z}_n$. For the case of one interval, $\mathcal{Z}_n= \langle \mathcal{T}_n \tilde{\mathcal{T}}_n \rangle $ is symmetric under cyclic permutations 
generated by the group $\mathbb{Z}_n$, hence only $\mathbb{Z}_n$-symmetric combinations of fields can enter the OPE 
$ \mathcal{T}_n  \tilde{\mathcal{T}}_n$.
In particular tensor products of primary fields of the single copy algebras (plus cyclic permutations) belong to this class of fields. 
But they are not the only ones: more primaries can be constructed from the linear combination of tensor products of primary and descendants fields in different copies (in the following sections we will give explicit examples in concrete models).

Note also that, generally speaking, in this enlarged CFT there exist degenerate fields, i.e., fields with the same scaling dimensions and this is not a condition under which Zamolodchikov formula (Section \ref{section-Zam}) is derived.
The obvious example would be to consider multiplets charged under permutation symmetry. However, as a consequence of the symmetry considerations above, just a combination of them enter the OPE \eqref{OPETTcnzn} and therefore the conformal blocks expansion, thus excluding the presence of this kind of degeneracies.
Still, we cannot exclude the presence of other degeneracies at higher order.
This possibility surely deserves more investigation. However, for our analysis this is not an issue, since the leading fusion channels we are going to consider do not show any degeneracy, so that the Zamolodchikov recursion formula holds

The fusion algebra of twist fields \eqref{OPETTcnzn} allows us to derive (following the exact same steps of Sec. \ref{section-CBE}) 
an expansion in conformal blocks. 
Making use of global conformal invariance, 
we can factorise the 4-point correlation function as (in the notations of \cite{CCT})
\begin{equation} \label{corr_functionTwist}
\langle \mathcal{T}_n (u_1) \tilde{ \mathcal{T}  }_n (v_1) \mathcal{T}_n (u_2) \tilde{\mathcal{T} }_n (v_2) \rangle = \left( \frac{(u_2-u_1)(v_2 - v_1)}{(v_1 - u_1) (v_2- u_2) (v_2 - u_1) (u_2 - v_1)} \right)^{4 \Delta_n} \mathcal{F}_n (x).
\end{equation}
The function $\mathcal{F}_n (x)$ can thus be expanded as
\begin{equation} \label{conformal_blocks}
\mathcal{F}_n (x) = \sum_{ \alpha } \left( C^{ \alpha }_{\mathcal{T}_n \tilde{\mathcal{T}}_n }\right)^2  F (x, nc, \Delta_{ \alpha} ,  \boldsymbol{\Delta}_n) F ( x ,nc, \bar{\Delta}_{\alpha} , \bar{\boldsymbol{\Delta}}_n),
\end{equation}
where $\boldsymbol{\Delta}_n \equiv \{ \Delta_{n }, \Delta_{n }, \Delta_{n }, \Delta_{n } \}$ and the structure constants $C^{\alpha}_{\mathcal{T}_n \tilde{\mathcal{T}}_n }$ can be related to the coefficients of the small $x$ expansion given in \cite{CCT}. The first terms have also been already computed (see \cite{CCT, Haedrick}).

\section{Zamolodchikov recursion formula} \label{section-Zam}

The computation of conformal blocks is an old problem in CFT. To this aim, one of the most powerful techniques is the Zamolodchikov recursion formula \cite{Zam} which turns 
out to be very useful in our case, because it provides an expansion where each term can be analytically continued to $n=1$.

The Zamolodchikov formula is an expansion in the elliptic variable 
\be
q(x)= e^{i \pi \tau(x)}, \quad \tau(x) = i\frac{K(1-x)}{K(x)},
\ee
where $K(x)$ is the complete elliptic integral of first kind and $x$ the usual four-point ratio \eqref{4pR}.
Clearly, small $q$ corresponds to small $x$, and the small $x$ expansion can be recast in terms of small $q$ expansion. 
Anyhow, it turned out that the expansion in $q(x)$ converges for finite $x$ much faster than the direct $x$ expansion \cite{GR}.

According to the Zamolodchikov formula (under the hypothesis of non-degenerate fields), the conformal block $F (x, \tilde{c},  \Delta_l , \boldsymbol{\Delta})$ 
satisfies the following recursion relation 
\begin{equation}  \label{Frecursion}
F (x, \tilde{c}, \Delta_l, \boldsymbol{\Delta}) = (16 q)^{\Delta_l - \frac{\tilde{c} -1}{24}} x^{ \frac{\tilde{c} -1}{24}} (1-x)^{ \frac{\tilde{c} -1}{24}} \theta_3 (\tau)^{\frac{\tilde{c}-1}{2} - 4 \sum_{i=1}^{4} \Delta_i} H (\tilde{c}, \Delta_l , \boldsymbol{\Delta}, q) ,
\end{equation} 
\begin{equation} 
H (\tilde{c}, \Delta_l , \boldsymbol{\Delta}, q)  = 1+ \sum_{m, n} (16 q)^{mn} \frac{R_{mn} (\tilde{c}, \boldsymbol{\Delta})}{\Delta_l - \Delta_{mn} (\tilde{c})} H (\tilde{c}, \Delta_{mn} + mn , \boldsymbol{\Delta}, q)  \label{Hrecursion},
\end{equation} 
where 
\begin{itemize}
\item $ \Delta_{mn} (\tilde{c})= \frac{\tilde{c} -1}{24} + \frac{(\beta_{+} m  + \beta_{-} n)^2 }{4}$,
\item $\beta_{\pm} = \frac{1}{\sqrt{24}} \left( (1 - \tilde{c})^{1/2}  \pm (25 - \tilde{c})^{1/2} \right)$,
\item $\theta_3$ is the Jacobi elliptic function,
\item $R_{mn}(\tilde{c}, \boldsymbol{\Delta}) = - \frac{1}{2} \prod'_{ab} \frac{1}{\lambda_{ab}} \prod_{p, q} (\lambda_1 + \lambda_2 - \frac{\lambda_{pq}}{2}) (-\lambda_1 + \lambda_2 - \frac{\lambda_{pq}}{2}) (\lambda_3 + \lambda_4 - \frac{\lambda_{pq}}{2}) (\lambda_3 - \lambda_4 - \frac{\lambda_{pq}}{2}) $,
\item $\lambda_{pq}= p \beta_{+} - q \beta_{-}$,
\item $\Delta_i = \frac{\tilde{c} -1}{24} + \lambda^2_i$,
\item the range of the indices run over
\begin{eqnarray*}
p &=& -m +1, m+3 , \cdots , m-3 , m-1,\\
q &=& - n +1 , -n +3, \cdots , n-3 , n-1, \\
a &=& -m+1, -m +2 , \cdots , m-1, m,\\
b &= & -n +1 , -n +2, \cdots , n-1, n,
\end{eqnarray*}
\item $\prod'_{ab}$ means $(a, b)  \neq (0, 0), (m, n)$.
\end{itemize}
The function $H$ can be given as an expansion in power of $q(x)$
\begin{equation} 
H (\tilde{c}, \Delta_l , \boldsymbol{\Delta}) = 1 + \sum_{k=1} h_k (\tilde{c}, \Delta_l, \boldsymbol{\Delta}) (16 q)^k. \label{Hsum}
\end{equation}
The first orders can be  explicitly written plugging (\ref{Hsum}) into (\ref{Hrecursion}), obtaining:
\begin{itemize}
\item[-] $k=1$
\begin{equation}
h_1 (\tilde{c}, \Delta_l , \boldsymbol{\Delta}) = \frac{R_{11} (\tilde{c}, \boldsymbol{\Delta}) }{\Delta_l - \Delta_{11}};
\end{equation}
\item[-]$k=2$
\begin{equation}
h_2 (\tilde{c}, \Delta_l , \boldsymbol{\Delta}) = 
\frac{R_{21} (\tilde{c}, \boldsymbol{\Delta}) }{\Delta_l - \Delta_{21} } + 
\frac{R_{12} (\tilde{c}, \boldsymbol{\Delta}) }{\Delta_l - \Delta_{12} } +
\frac{R_{11}^2 (\tilde{c}, \boldsymbol{\Delta}) }{\Delta_l - \Delta_{11} };
\end{equation}
\item[-] $k=3$
\begin{multline} \label{h3}
 h_3 (\tilde{c}, \Delta_l, \boldsymbol{\Delta}) =
\frac{R_{31} (\tilde{c}, \boldsymbol{\Delta}) }{\Delta_l - \Delta_{31}} + 
\frac{R_{13} (\tilde{c}, \boldsymbol{\Delta}) }{\Delta_l - \Delta_{13} } + \\
 +    \frac{R_{21} (\tilde{c}, \boldsymbol{\Delta}) }{\Delta_l - \Delta_{21} }   \frac{R_{11} (\tilde{c}, \boldsymbol{\Delta}) }{\Delta_{21} +2 - \Delta_{11} }  + \frac{R_{12} (\tilde{c}, \boldsymbol{\Delta}) }{\Delta_l - \Delta_{12} }   \frac{R_{11} (\tilde{c}, \boldsymbol{\Delta}) }{\Delta_{12} + 2 - \Delta_{11}} + \\
 + \frac{R_{11} (\tilde{c}, \boldsymbol{\Delta})   }{\Delta_l - \Delta_{11}} \left(  \frac{R_{21} (\tilde{c}, \boldsymbol{\Delta}) }{\Delta_{11} +1 - \Delta_{21}} + 
\frac{R_{12} (\tilde{c}, \boldsymbol{\Delta}) }{\Delta_{11} + 1 - \Delta_{12}} +
R_{11}^2 (\tilde{c}, \boldsymbol{\Delta})  \right).
\end{multline}
\end{itemize}
However, as shown in \ref{appendix0}, for correlation functions of fields with the same conformal dimensions (as in the case we are interested in) only the even powers appear in Eq. \eqref{Hsum}, which therefore takes the form
\begin{equation}
H (\tilde{c}, \Delta_l , \boldsymbol{\Delta}) = 1 + \sum_{k=1} h_{2k} (\tilde{c}, \Delta_l, \boldsymbol{\Delta}) (16 q)^{2k}.
\end{equation}

Note that in Eq. \eqref{Hrecursion} the dimensions of the fusion channels appear in the denominator of Zamolodchikov 
formula so that singularities could be present for
\begin{equation}
\Delta_l = \Delta_{mn} (\tilde{c}).
\end{equation}
In particular, when $\tilde{c}=1$, the denominator vanishes also for the identity channel
(since $\Delta_l = \Delta_{11} (\tilde{c}=1) =0$).
Moreover, still in the case $\tilde{c}= 1$, all the factors $R_{mn}$ with $(m \cdot n) \geq 2$ show null denominator
and hence all $h_{k \geq2}$ do the same.
We encounter this problem in both the examples considered here: 
for the function $\mathcal{F}_2 (x)$ in the Ising model ($\tilde{c} = n c =1$, if $n=2$) 
and in the limit $n\to1$ for the compact boson ($\tilde{c} =nc = 1$ for $n=1$). 
However, we will show that in these two cases the limit $\tilde{c} \to 1$ exists, so that there is no problematic issue.

\subsection{Truncations of Zamolodchikov recursion formula} \label{approx}

In the previous section we discussed how the conformal block technique can be generalised  in order to compute the function 
$\mathcal{F}_n(x)$ for the 4-point correlation function of twist fields, as already pointed out in \cite{GR}.
Eq. \eqref{conformal_blocks} is in fact a rewriting of the entire correlation function in terms of conformal blocks as building blocks. 
However, it is still unknown and probably impossible to resum the entire series, even for the 
easiest models. 
In the practical world, we are just able to truncate this formula, but there are several levels of truncations that can play a role, 
as we are going to discuss in the following.

The first one is a truncation of the conformal block expansion, Eq. (\ref{conformal_blocks}). 
In a general model, the sum over the fusion channels is actually a series.
Moreover the expansion involves more and more channels as 
the replica label $n$ increases (as it should be clear from the structure of the generalised  OPE of twist fields, cf. Eq. \eqref{OPETT}). 
Therefore, when interested to generic $n$, we must truncate this sum to the first leading terms, depending on the accuracy we
wish to reach (in the following we will see how to order the contributions of the 
different channels from the most to the less relevant).
In \cite{GR}, for the Ising model,  a truncation to the first two leading channels was considered, and it provided a 
good approximation of $\mathcal{F}_2$ and $\mathcal{F}_3$ only. 
In the following, we are going to keep more terms in this expansion discussing how the final result for the entanglement entropy at $n=1$
(which requires the knowledge of $\mathcal{F}_n$ as a function of $n$) may be improved.

The second truncation is in the order in the recursion formula, i.e. in Eq. (\ref{Hrecursion}) we must fix a $\bar{k} \in \mathbb{N}$ s. t. 
\begin{equation}
H (\Delta_l, \boldsymbol{\Delta}) \sim 1 + \sum_{k = 1}^{\bar{k}} h_{2k }(16 q(x))^{2k}.
\label{trunc2}
\end{equation}
Also in this case, the most important issue is to understand whether the first few terms in this series are enough to 
get a good approximation. 
For example, for the Ising model, it turned out \cite{GR} that a good approximation is obtained already at the 
zeroth-order $H(\Delta_l, \boldsymbol{\Delta}) \sim 1$.  
Here we will show that by keeping more terms in \eqref{trunc2} it is possible to practically get convergence of the function, 
so that this truncation is minimally affecting the final result.

\section{An overview of some exact results for the entanglement entropy of two disjoint intervals in CFT}
\label{overview}

In the following sections we will apply the Zamolodchikov recursion formula to entanglement entropies of the 
critical Ising model (aka $\mathcal{M}_{3}$ minimal model) and of the massless compact boson, which are CFTs with central 
charge equal to $1/2$ and $1$ respectively. 
In this section we report the known exact results for the moments of the reduced density matrices for these two models since we will 
repeatedly compare our truncated expressions with them. 

For the free boson compactified on a circle of radius $R$, the scaling function ${\cal F}_n(x)$ for general integers has been 
calculated in Ref. \cite{CCT-1} (generalising the result at $n=2$ in \cite{fps-09}) and it reads
\begin{equation}
{\cal F}_n(x)=
\frac{\Theta\big(0|\eta\Gamma\big)\,\Theta\big(0|\Gamma/\eta\big)}{
[\Theta\big(0|\Gamma\big)]^2}\,,
\label{Fnb}
\end{equation}
where $\eta=2R^2$, $\Gamma$ is an $(n-1)\times(n-1)$ matrix with elements \cite{CCT-1}
\be
\Gamma_{rs} =  
\frac{2i}{n} \sum_{k\,=\,1}^{n-1} 
\sin\left(\pi\frac{k}{n}\right)\beta_{k/n}\cos\left[2\pi\frac{k}{n}(r-s)
\right]\,, 
\label{Gammadef}
\ee 
\be
\beta_y=\frac{\, _2 F_1(y,1-y;1;1-x)}{\, _2 F_1(y,1-y;1;x)}\,,
\label{betadef}
\ee
and $\Theta$ is the Riemann-Siegel theta function
\begin{equation}
\label{theta Riemann def}
\Theta(0|\Gamma)\,\equiv\,
\sum_{m \in \mathbf{Z}^{n-1}}
\exp\big[\,i\pi\,m^{\rm t}\cdot \Gamma \cdot m\big]\,.
\end{equation}
For the Ising model, the scaling function ${\cal F}_n (x)$ is \cite{CCT}
\begin{equation}
{\cal F}_n (x)= 
\frac{1}{2^{n-1}\Theta({\bf 0}|\Gamma)} \sum_{\bm{\varepsilon,\delta}}
\left| \Theta\bigg[\begin{array}{c} \bm{\varepsilon} \\ \bm{\delta}  \end{array}\bigg] ({\bf 0}|\Gamma)\right|\,.
\label{Fni}
\end{equation}
Here $\Theta$ is the Riemann theta function with characteristic defined as
\begin{equation}
\label{def theta}
\Theta
\bigg[\begin{array}{c} \bm{\varepsilon} \\ \bm{\delta}  \end{array}\bigg] ({\bf z} |\Gamma)
\,\equiv\,
\sum_{{\bf m} \in \mathbf{Z}^{n-1}} \exp\Big[i \pi ({\bf m+{\bm \varepsilon}})^{{\rm t}}\cdot \Gamma\cdot ({\bf m+\bm{\varepsilon}})
+2\pi i\,({\bf m+\bm{\varepsilon}})^{{\rm t}}\cdot ({\bf z+\bm{\delta}})\Big]\,,
\end{equation}
where ${\bf z} \in \mathbf{C}^{n-1}$ and $\Gamma$ is the same as in Eq. (\ref{Gammadef}).
$\bm{\varepsilon}, \bm{\delta} $ are vector with entries  $0$ and $1/2$.
The sum in $(\bm{\varepsilon,\delta})$ in Eq. (\ref{Fni}) is intended over all the $2^{n-1}$ vectors 
${\bm \varepsilon}$ and ${\bm \delta}$ with these entries. 
This result generalises the one for $n=2$ in \cite{ATC-Ising}.

Finally, the {\it universal  scaling function} $F_{\rm vN}(x)$ for the Von Neumman is usually defined as
\be
F_{\rm vN}(x)\equiv S_{A_1}+S_{A_2}-S_{A_1\cup A_2}-\frac{c}3 \ln (1-x)\,,
\ee
where $A_1$ and $A_2$ are the two intervals we are focusing on. Notice that the combination of entropy in the rhs is nothing 
but the mutual information which indeed is scale invariant.

\section{Ising model}  \label{section-Ising}

In this section we apply the machinery  of the conformal blocks expansion and the Zamolodchikov recursion formula to the minimal model 
$\mathcal{M}_{3}$, corresponding to the CFT describing the critical Ising model. 

\subsection{OPE of twist fields}

For the $\mathcal{M}_{3}$ CFT, the mother theory contains only a finite number of primary fields with natural working  basis
\be
\mathbb{I}, \sigma , \epsilon,
\ee
with $\mathbb{I}$ the identity, $\sigma$ the spin operator, and $\epsilon$ the energy density operator with dimensions 
$\Delta_\mathbb{I}=0$, $\Delta_\sigma=1/16$, and $\Delta_\epsilon=1/2$. However, since we now consider $n$ decoupled copies of the theory, the associated central charge is $\tilde{c}= n c \geq 1$ (if $n \geq 2$), therefore, as argued in \cite{cardy-86}, the number of primaries fields is in principle infinite (even if it may be reduced when taking into account the $\mathbb{Z}_n$ symmetry due to the boundary conditions connecting the different copies).

The OPE of $\mathcal{T}  _n \tilde{ \mathcal{T}  }_n$ takes the general form \eqref{OPETTcnzn}, which for the Ising case 
reduces to
\begin{equation}
\mathcal{T}_n \tilde{ \mathcal{T} }_n =1+ ([\sigma_i \sigma_j ] + \; perm) + ([\epsilon_i \epsilon_j] + \; perm) + ([\sigma_i \sigma_j \epsilon_k] + \; perm) + \cdots
\end{equation}
where \emph{perm} stands for all possible permutations of the indices from 1 to $n$.
In this notation,  the insertion of an operator in the family of the identity is implicit each time a given sheet is not explicitly  
indicated (see explicit examples below).
 
Since as we increase the number of sheets, there are more and more choices of operators, the families that must be considered 
depend on $n$. For example, for the lowest values of $n$, we have
\begin{itemize}
\item $n=2$:
\be \label{classificatio-n2}
\mathbb{I}_1 \mathbb{I}_2, \; \sigma_1 \sigma_2 , \; \epsilon_1 \epsilon_2.
\ee
Terms with only a single copy of any fields (e.g. $(\sigma_1 \mathbb{I}_2)$ , $(\epsilon_1 \mathbb{I}_2)$) are not present, 
as already stressed in \cite{GR}; also the term $(\sigma_1 \epsilon_2)$ is not there by symmetry. In this case it has been shown \cite{OPE-F2} that these families complete the OPE.
\item $n=3$:
\be \label{n3}
\mathbb{I}_1 \mathbb{I}_2 \mathbb{I}_3, \; \sigma_1 \sigma_2 \mathbb{I}_3 , \; \epsilon_1 \epsilon_2 \mathbb{I}_3 , \; \sigma_1 \sigma_2 \epsilon_3,
\; (L_{-1} \sigma)_1 \sigma_2 \mathbb{I} - \sigma_1 (L_{-1} \sigma)_2 \mathbb{I},
\ee
and permutations. Terms like $\sigma_1 \sigma_2 \sigma_3$ and $\epsilon_1 \epsilon_2 \epsilon_3$ vanish (due to the vanishing structure constants $C_{\sigma\sigma}^{\sigma}$ and $C_{\epsilon \epsilon}^{\epsilon}$\cite{DiFrancesco}). Note that the last example in Eq. \eqref{n3} is still a primary operator according to definition \eqref{primary_tot} but is not  in the form of a tensor product: in principle the associated OPE coefficient could be computed using the generalised formula in \cite{cz-13} but the calculation is more involved, therefore we do not include it in what follows. Other primaries of this type may in principle occur. All the other terms in Eq. \eqref{n3} will be included in our approximation. 
\end{itemize}
In the following we are going to denote as 
\be
C_{k,l} (n),
\ee
the coefficient of the generic term
\begin{equation} \label{klterm}
(\sigma_1 \cdots \sigma_k \epsilon_{k+1} \cdots \epsilon_{k+l} \mathbb{I}_{l+1} \cdots \mathbb{I}_{n} + perm),
\end{equation}
in the expansion in conformal blocks.
It can be related to the coefficient $s_{k, l}(n)$ entering the small $x$ expansion \cite{CCT} as
\begin{equation} \label{x-order}
 \left(  \frac{x}{4n^2}  \right)^{ 2 (k \Delta_{\sigma} + l \Delta_{\epsilon} ) } s_{ k, l} (n).
\end{equation}
In particular, one can show that it holds
\begin{equation} \label{x-order2}
C_{k, l} (n)^2 = \left( \frac{1}{4n^2}  \right)^{2(k \Delta_{\sigma}  + l \Delta_{\epsilon}) } s_{k, l} (n).
\end{equation}
Eq. \eqref{x-order} also provides a criterion to order the different fusion channels  from the most to the less relevant ones, by 
looking to the order at which they enter in the 4-point correlation function in the small $x$ expansion.

\subsection{The explicit results from recursion formula and comparison with the exact ones}

In this section we explicitly build the universal function $\mathcal{F}_n (x)$ for the ${\cal M}_3$ minimal model for various $n$
and at several different orders in the truncation of Zamolodchikov formula. 
We compare our results with the exact function $\mathcal{F}_n (x)$ for increasing values of $n$.
We also analytically obtain a truncation for the Von Neumann entropy scaling function $F_{\rm vN} (x)$  via replica trick 
and compare it with the very accurate results from numerical simulation in \cite{ATC-Ising}.

\paragraph{$0$-th order in the Zamolodchikov recursion formula.}

We start by truncating  the Zamolodchikov recursion formula to the $0$-th order (corresponding to $H \sim 1$ in \cite{GR}) 
and we proceed by including more and more terms in the OPE expansion of twist fields to reach a reasonable approximation  
of the function $\mathcal{F}_n (x)$, for a given $n$.
As already stressed, the number of terms expected from the OPE is increasing quickly with $n$.
Thus a good approximation requires more and more terms as $n$ increases. 

\begin{figure}[t]
       \centering
        \subfigure
       {\includegraphics[scale=0.3]{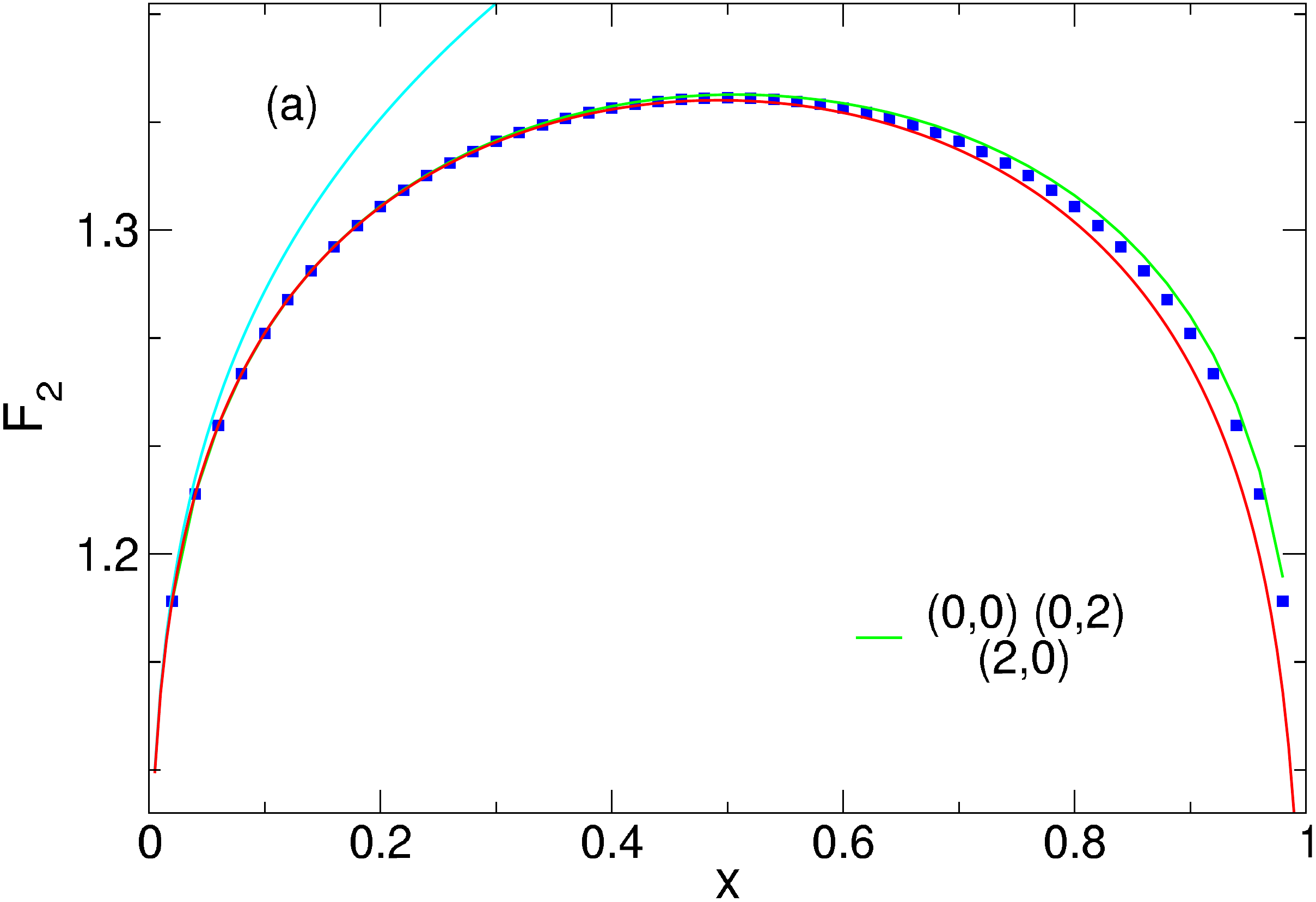} } 
               \subfigure
       {\includegraphics[scale=0.3]{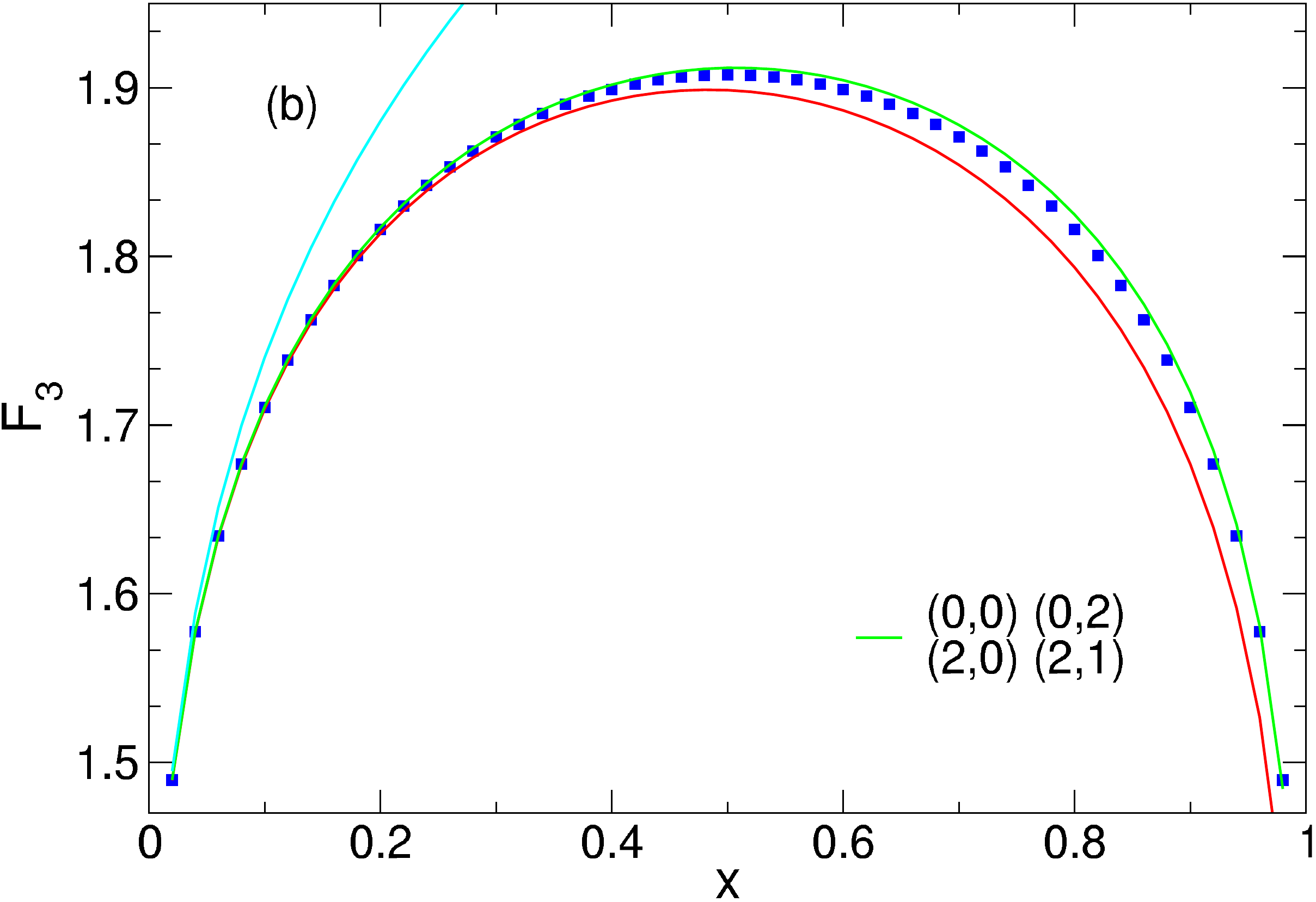}}  
               \subfigure
       {\includegraphics[scale=0.3]{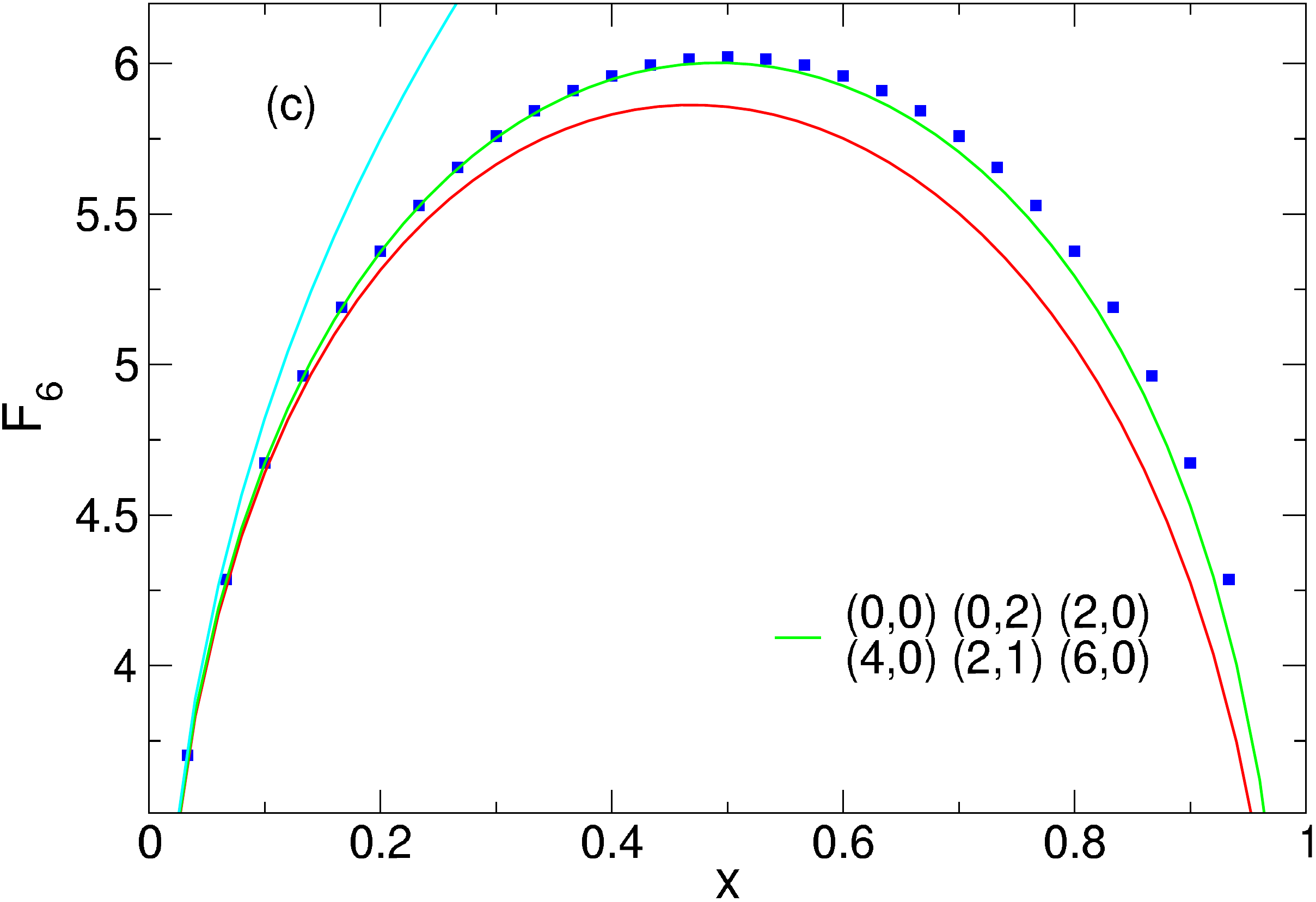}}  
           \subfigure
     {\includegraphics[scale=0.3]{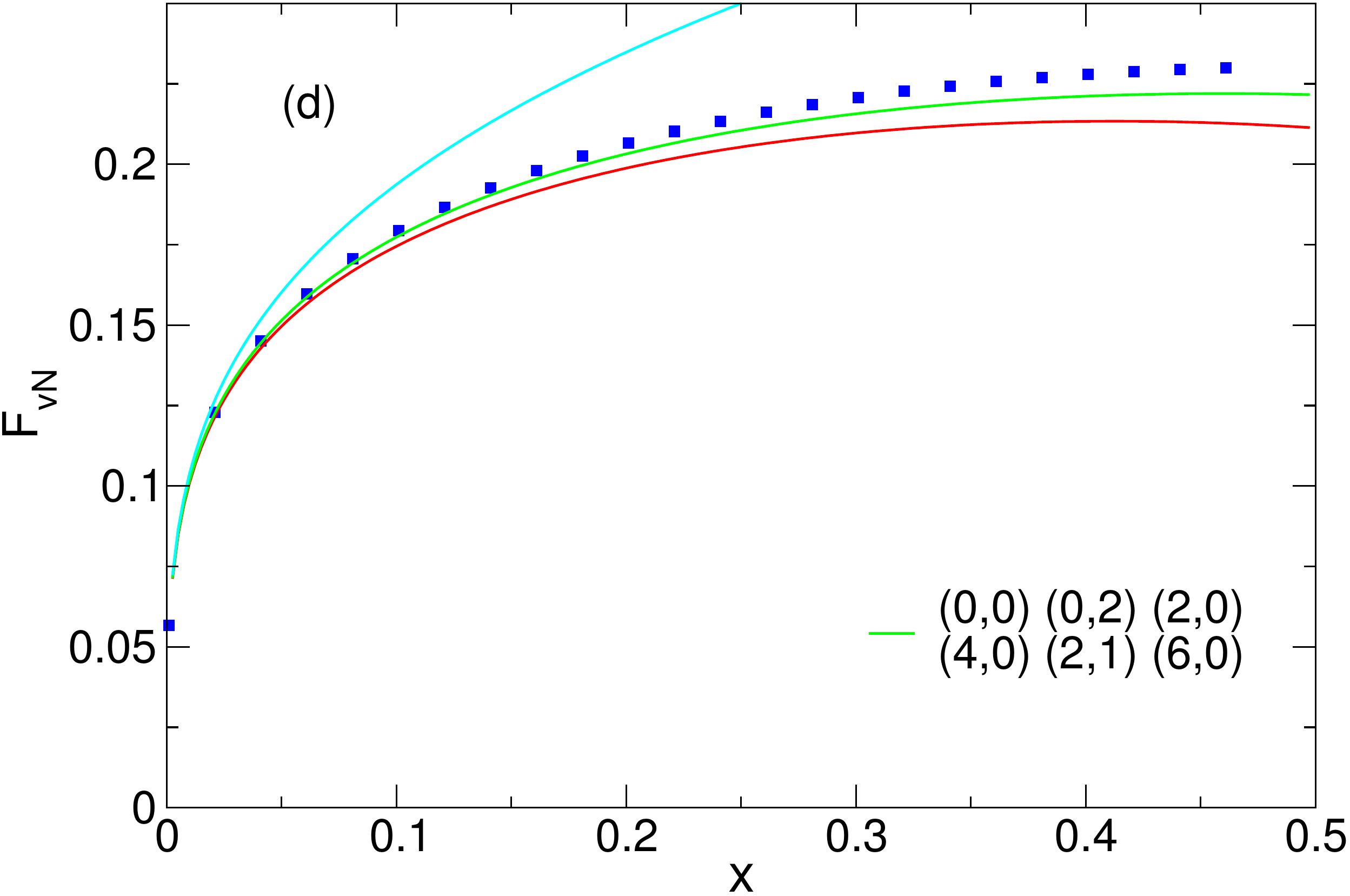}} 
        \caption{The best approximation we derived for $\mathcal{F}_n (x)$ ($n=2,3, 6$) and for the entanglement entropy $F_{\rm vN} (x)$ in the Ising model, still at the 0-th order in the Zamolodchikov formula. The dots represent the exact functions. The red line is the curve derived with the approximation in \cite{GR}. The green curve is our approximation (the fusion channels included in the OPE of twist fields are listed with $(k, l)$ denoting the inclusion of the fusion channel $ [\sigma_1 \cdots \sigma_k \epsilon_1 \cdots \epsilon_l \mathbb{I}_{k+l+1} \cdots \mathbb{I}_n ]$ and all its permutations). The cyan curve is the expansion in power of $x$ derived in \cite{CCT}.}
         \label{Fn_Ising_best}
  \end{figure}

In Figure \ref{Fn_Ising_best} (a), (b) and (c), we report the result for the zeroth order (green curves) for $n=2,3,6$.
We compare this zeroth order truncation (including several channels in the OPE) with 
the known exact results, with the truncation of \cite{GR} (which includes the first two channels only), 
and with the small $x$ expansion of \cite{CCT}. 
It is evident that including more channels in the OPE considerably improves the approximation which is extremely close to the 
exact result. 
In the figures we denoted by $(k, l)$ the truncation with the inclusion of the fusion channel 
$ [\sigma_1 \cdots \sigma_k \epsilon_1 \cdots \epsilon_l \mathbb{I}_{k+l+1} \cdots \mathbb{I}_n ]$ (and all its permutations). 

It is evident that for some values of $x$, our approximation gives a curve which is slightly larger than the exact result.
We will see that the curve will be moved downward by the inclusion of higher terms in the recursion formula.

The von Neumann entropy can be obtained at a given order by analytic continuation.
The first few terms leads to the truncation for the scaling function 
\begin{multline}
F_{\rm vN}^{(\text{0-th})} (x) =   \theta_3^{-\frac{1}{2}} (q) \left( \frac{x (1-x)}{16 q}  \right)^{-\frac{1}{24} } \times 
\\ \times \left[ - \frac{5}{6} \ln \theta_3 (q) + \frac{1}{24}  \ln \left( \frac{x(1-x)}{16 q} \right)  + s'_{2, 0} (1) (4q)^{\frac{1}{4}}  + s'_{4, 0} (1) (4q)^{\frac{1}{2}} + \cdots \right].
\label{Fvn}
\end{multline}  
The coefficients $s'_{k, l} (1)$ are calculated in \ref{appendix} by analytic continuation.
All the other fusion channels give an additive contribution implicit in the dots above.
Note that, even if a finite number of terms in the Zamolodchikov expansion may exactly reproduce 
$\mathcal{F}_n (x)$ for finite $n$, the same is not true for $F_{\rm vN} (x)$, since an infinite number of terms contributes to the 
analytic continuation. 
 
In panel (d) of Figure \ref{Fn_Ising_best} we report the von Neumann entropy scaling function $F_{\rm vN}(x)$ \eqref{Fvn} and 
we compare it with the results from numerical simulations in \cite{ATC-Ising} (we only report data for $x<0.5$, the other half is better 
reproduced exploiting the symmetry $x\to1-x$). 
We notice that the agreement of the truncation with the numerical data is reasonable, but not as good as those at finite $n$.     
In fact, although we included a number of terms reproducing well $\mathcal{F}_n (x)$ up to $n=6$, 
the truncation for the entanglement entropy deviate considerably from the exact numerical data, but it is still 
a much better approximation than the one considered in \cite{GR}.

\paragraph{$M$-th order in the Zamolodchikov recursion formula.}

\begin{figure}[t] 
       \centering%
        \subfigure
       {\includegraphics[scale=0.3]{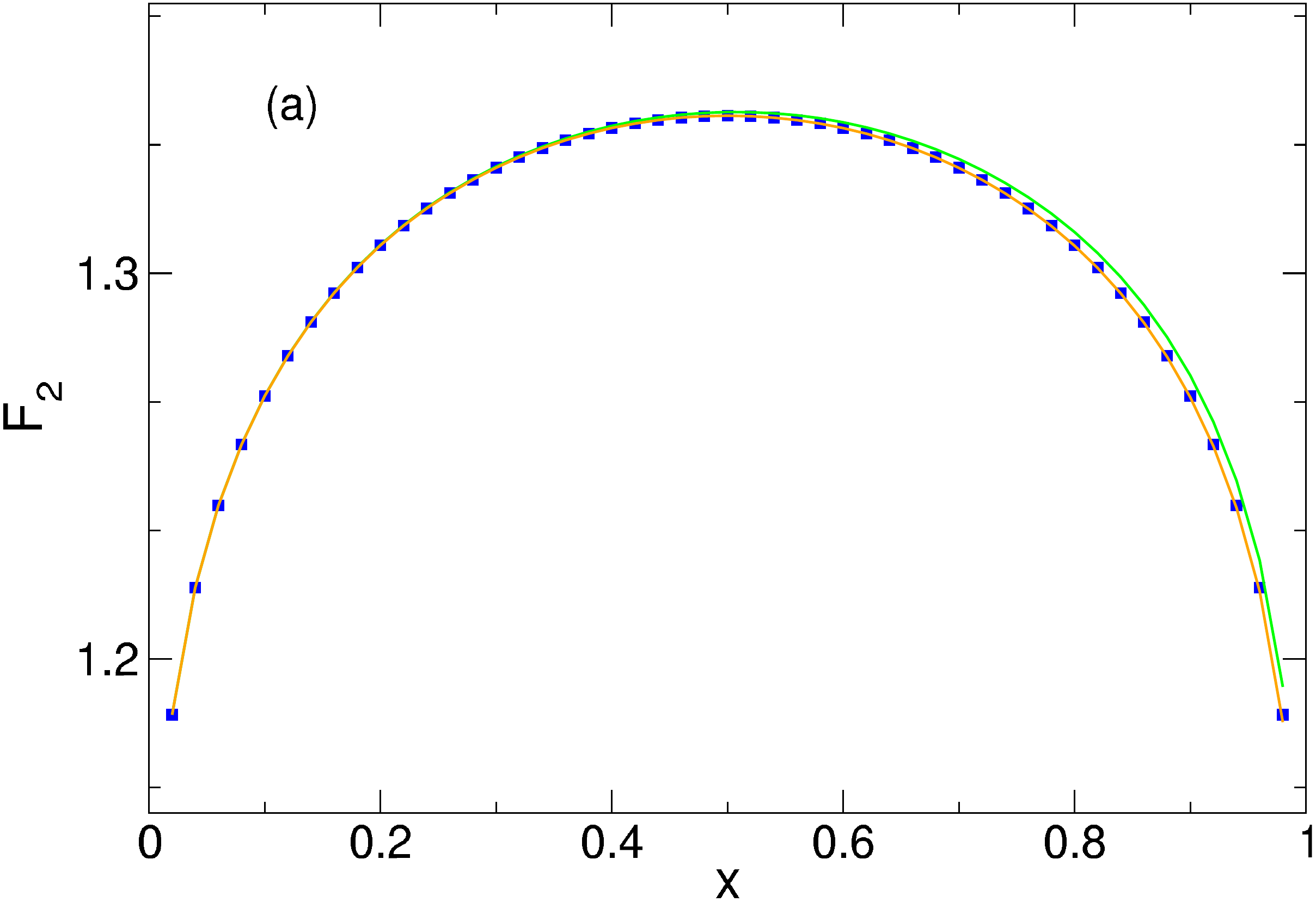}}
               \subfigure
       {\includegraphics[scale=0.3]{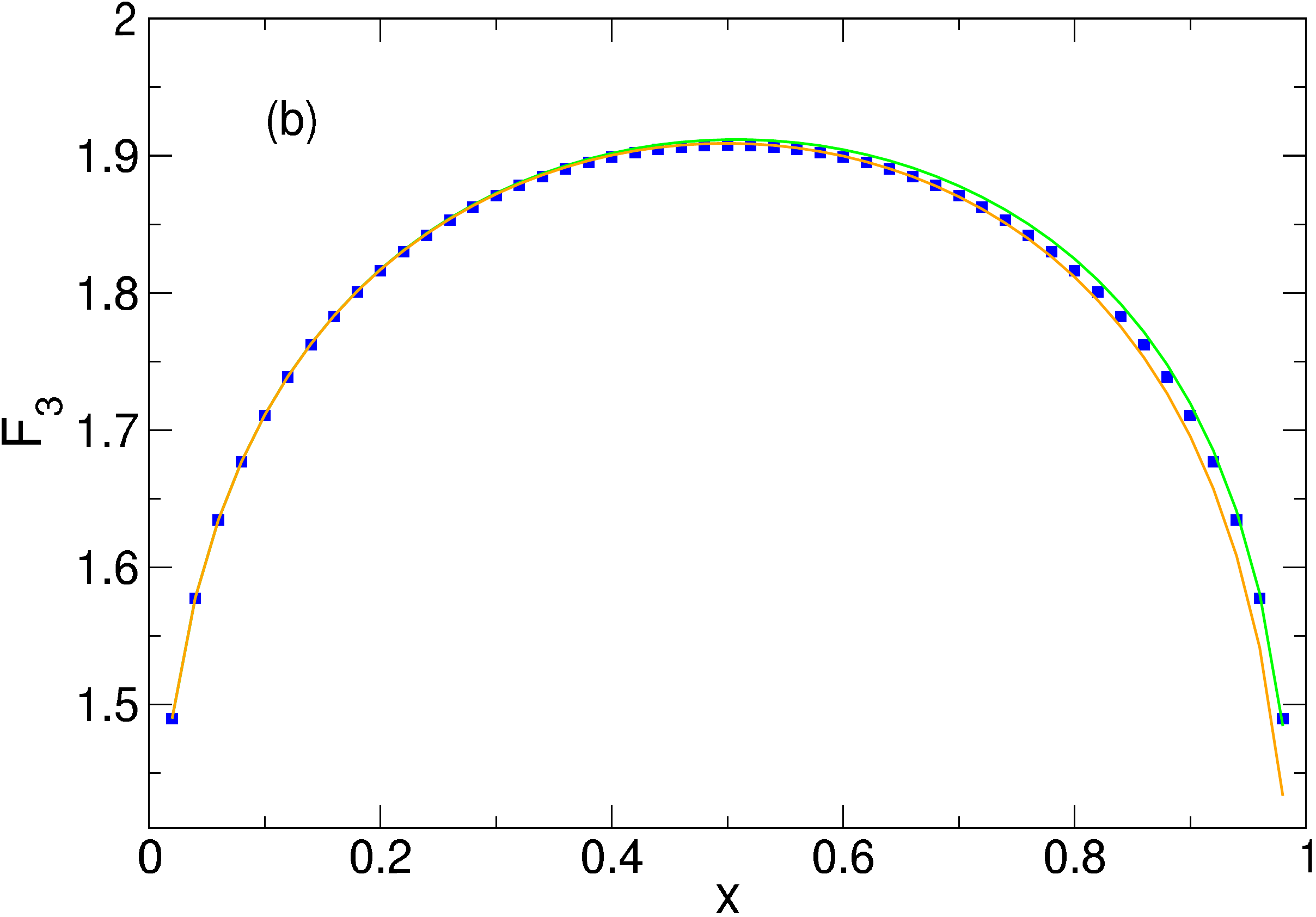}}
               \subfigure
       {\includegraphics[scale=0.3]{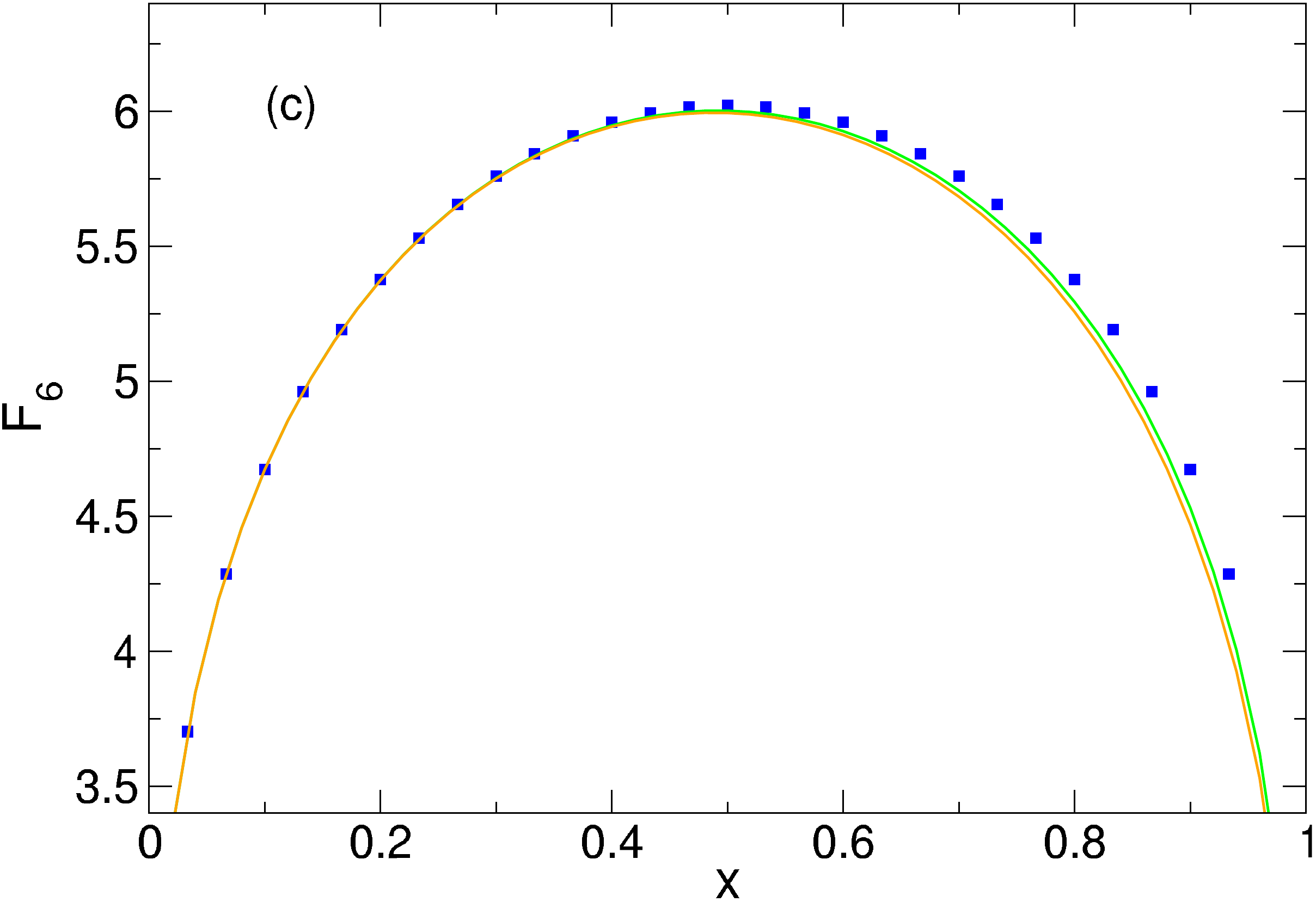}}      
                    \subfigure
       {\includegraphics[scale=0.3]{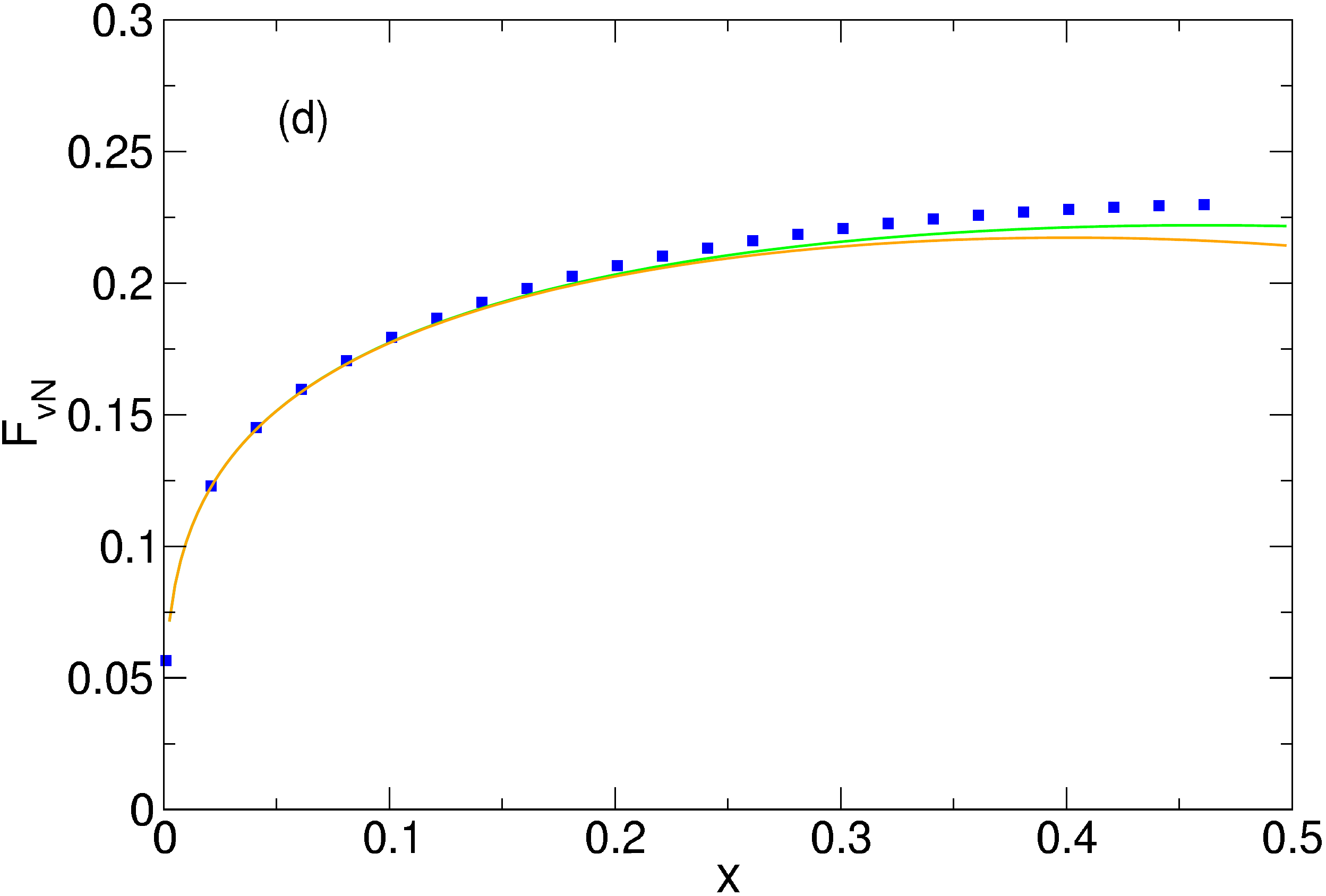}}    
        \caption{The best approximation we derived for $\mathcal{F}_n (x)$ ($n=2, 3, 6$) and the von Neumann entropy $F_{\rm vN} (x)$ in the Ising model, by including further terms in the Zamolodchikov formula. The dots represent the exact functions. The green curve is our approximation at the 0-th order. The orange curve is the approximation at the $2$-nd order. The fusion channels included in the OPE of twist fields are the same as in Figure \ref{Fn_Ising_best}.}
        \label{Fn_Ising_convergence}
  \end{figure}

We now discuss how the approximation improves by taking into account more iterations in the Zamolodchikov recursion formula. 
It turns out that the Zamolodchikov series converges extremely fast for the functions $\mathcal{F}_n (x)$: 
the truncation to the first 2 orders in $q(x)$ (namely $H \sim 1 + h_2 q(x)^2$) is practically indistinguishable from the function we obtain 
by summing up numerically the whole series. 
The results of this improved truncation are shown in Figure \ref{Fn_Ising_convergence} (a), (b) and (c) for $n=2,3,6$ respectively. 
In the figure, the results are compared to those computed at the $0$-th order in the recursion formula and to the exact results.  
It is evident that the already very accurate truncation at zeroth order is further improved by the iteration of the recursion formula for $n=2, 3$.
In particular for $n=2$ the approximated result is indistinguishable from the exact one: this fact does not come unexpected because this is the only case where we know the OPE of twist fields to be complete (cf. Eq. \eqref{classificatio-n2}) and the recursion formula has converged.
For $n=6$ instead the agreement is imperceptibly worse.

We can also improve the truncation for the von Neumann entropy. 
The scaling function $F_{\rm vN} (x)$, as derived via replica trick, at the $2$-nd order in the recursion formula takes the form
\begin{multline}
F_{\rm vN}^{(\text{2-nd})} (x) =   \theta_3^{-\frac{1}{2}} (q) \left( \frac{x (1-x)}{16 q}  \right)^{-\frac{1}{24}} \times  \\ \times \left[ \left(- \frac{5}{6} \ln \theta_3 (q) + \frac{1}{24}  \ln \left( \frac{x(1-x)}{16 q} \right) \right)(1+ 2 h_2^{(0,0)} (1)(16 q)^2)  + \right. \\
+ 2 h_2^{(0, 0)}{}'(1)\, (16 \, q)^2  
  + s'_{2, 0} (1) (4q)^{\frac{1}{4}} \left(1+ 2 h_2^{(2, 0)} (1)\, (16 \, q)^2 \right) +\\
\left.  + s'_{4, 0} (1) (4q)^{\frac{1}{2}} \left(1 + 2 h_2^{(4,0)}(1) \, (16 \, q)^2 \right) + \cdots \right],
\end{multline} 
where the coefficients $h_2^{(k, l)} (n)$ are a shortcut for the coefficients of the expansion in Eq. \eqref{Hsum} 
\begin{equation}
h_2^{(k, l)} (n) \equiv h_2 (n c , \Delta_{(k, l)}, \boldsymbol{\Delta}_n),
\end{equation}
for a given conformal family (identified by $(k, l)$, with $\Delta_{(k,l)}$ its conformal dimension), which takes the simple form
\begin{equation}
h_2^{(k, l)} (n) =
\frac{(-n c  + ( nc - 32 \Delta_n)^2 + 
 2 \Delta_{(k, l)} (1 + n c  - 32 \Delta_n) (5 +  n c - 32 \Delta_n))}{(512 ( n c + 
   2 \Delta_{(k, l)} (-5 + 8 \Delta_{(k, l)}+  n c )))} 
\end{equation}
This higher order truncation is shown in panel (d) of Figure \ref{Fn_Ising_convergence}: we notice that it does not provide an improvement of the zeroth-order result of $F_{\rm vN}(x)$ for large values of $x$.


\section{Compact boson}  \label{section-CB}

In this section we apply the  Zamolodchikov recursion formula to the truncation of the entanglement entropies in the 
conformal field theory of a free massless boson compactified on a circle of radius $R$, which has 
central charge $c=1$.

\subsection{OPE of twist fields}


As for the Ising model, the starting point of our analysis  is the OPE of twist fields which is always of the form \eqref{OPETTcnzn}. 
The main difference with respect to the Ising model is that, while for the latter a basis of local field in the single copy theory is 
given by a set of three fields only ($\mathbb{I}, \sigma, \epsilon$), for the compact boson we have an infinite set already in the 
mother theory.

The most relevant fields we consider are the derivative operators 
\be
\partial_z \varphi (z), \qquad {\rm and}\qquad \partial_{\bar{z}} \bar{\varphi} (\bar{z}),
\ee 
whose conformal weights are $(1, 0)$ and $(0, 1)$ respectively, and the \emph{vertex operators}, which are uniquely identified by a pair of integers ($m, n$)
\begin{equation}
V_{(m,n)} \equiv  \; : \exp ( i \alpha_{m, n} \varphi (z)  + i \bar{\alpha}_{m, n} \bar{\varphi} (\bar{z})  )   :,
\label{vertex}
\end{equation}
where $\alpha_{m, n}$ and $ \bar{ \alpha }_{m, n}$ are the holomorphic and 
antiholomorphic \emph{charges} 
\begin{equation}
{ \alpha }_{m, n} = \left( \frac{m}{ \sqrt{2 \eta} }  + n \sqrt{\frac{\eta}{2}}  \right) ,
\quad   \bar{\alpha  }_{m,n} = \left( \frac{m}{ \sqrt{2 \eta} }  - n \sqrt{\frac{\eta}{2}}  \right)    .
\end{equation} 
where $\eta=2 R^2$ is a function of the compactification radius $R$.
They are associated to the vertex operators of conformal dimensions
\begin{equation}
h_{m, n} = \alpha_{m, n}^2 /2  \qquad  \bar{h}_{m, n} = \bar{\alpha}_{n, m}^2 /2.
\end{equation}

Of course for the replicated theory, the primary fields with respect to the total Virasoro algebra are infinitely many.
However, many of them do not appear in the fusion algebra of $\mathcal{T}_n  \tilde{\mathcal{T}}_n$. 
For example, for primaries constructed as tensor product of vertex operators on each copy, the structure constants are proportional to the correlator \cite{CCT}
\begin{equation}
C_{\{ m_j , n_j\}  }  \propto  \langle \;  \prod_j V_{(m_j, n_j)} (e^{2 \pi i j /n}) \; \rangle_{\mathbb{C}},
\end{equation}
which due to the \emph{neutrality condition} \cite{DiFrancesco} vanishes unless 
\begin{equation} \label{neutrality}
\sum_i \alpha_{m_i , n_i} = 0  \qquad \sum_i \bar{\alpha}_{n_i , m_i} =0.
\end{equation}

In our analysis we will not consider the contribution to $\mathcal{F}_n(x)$ from the two point function of the derivative operators 
because of their complicated analytic structure. Indeed, since the derivative operator has non zero conformal spin $s= 1$,
as shown in \cite{CCT}, its contribution vanishes unless $4 s/n \in \mathbb{Z}$. 
As a consequence, also the analytic continuation at $n =1$ is highly non-trivial.

The non-vanishing primary terms that we consider are of the form 
\begin{equation}
(  \underbrace{V_{ ( m, 0 ) } \cdots V_{ ( m, 0 ) }}_{k} \underbrace{ V_{ ( -m, 0) } \cdots V_{ ( -m, 0 ) }}_{k} \, + \,  perm  ),
\end{equation} 
with $k \leq n/2$.
At the leading order in the small $x$ expansion of the conformal block, they contribute as $x^{k m^2 / 2\eta}$, meaning that their contribution to the 4-point correlation function is of order $x^{k m^2 / \eta}$.
%
Similarly, operators of the form
\begin{equation}
(  \underbrace{V_{ ( 0, n) } \cdots V_{ ( 0, n ) }}_{l} \underbrace{ V_{ ( 0, -n) } \cdots V_{ ( 0, -n ) }}_{l} \, + \,  perm  ),
\end{equation} 
contribute in the 4-point correlation function as $x^{l n^2 \eta }$.
Consequently,  the most general non vanishing combination of vertex operators in the small $x$ expansion
gives rise to terms of order
\begin{equation}
x^{k \frac{m^2}{\eta} + l n^2 \eta  }.
\end{equation} 
In the present case then, the relevance of the different  contributions depend not  only on the number of copies $n$, 
but also on the parameter $\eta$ and consequently it is less obvious how to order them.
The leading contribution either comes from the fusion channel
\begin{equation}
([ V_{(1, 0)} V_{(-1, 0)} ]\, + \, perm),
\end{equation}
if $\eta <1$, or from 
\begin{equation}
([ V_{(0, 1)} V_{(0,-1)} ]\, + \, perm),
\end{equation}
if $\eta >1$.
Since there is a symmetry $\eta\to1/\eta$ \cite{DiFrancesco}, we continue by 
discussing only $\eta <1$,  for which a next-to-leading term is
\begin{equation}
([ V_{(1, 0)} V_{(-1, 0)}  V_{(1, 0)} V_{(-1, 0)}]\, + \, perm).
\end{equation}
For later convenience we define the coefficient of terms of the form
\begin{equation} \label{cb-fusionchannel}
(  \underbrace{V_{ ( p, 0 ) } \cdots V_{ ( p, 0 ) }}_{k} \underbrace{ V_{ ( -p, 0) } \cdots V_{ ( -p, 0 ) }}_{k} \,   \underbrace{V_{ ( 0, q) } \cdots V_{ ( 0, q ) }}_{l} \underbrace{ V_{ ( 0, -q) } \cdots V_{ ( 0, -q ) }}_{l} \, + \,  perm  ),
\end{equation} 
as
\begin{equation}
C_{k, l}^{(p, q)} (n),
\end{equation}
which turns out to be related to the coefficients of the small $x$ expansion $s_{2k,2l}^{(p, q)} (n)$ (introduced in analogy to the ones for the Ising model, cfr. Eq. \eqref{x-order} and Eq. \eqref{x-order2})
\begin{equation}
C_{k, l}^{(p, q)} (n)^{2} = \left( \frac{1}{4  n^2}  \right)^{\frac{k}{\eta} + l \eta } s_{2k,2l}^{(p, q)} (n).
\end{equation}
The inclusion of such a fusion channel is denoted in the figures by $(p, q; k, l)$.

\subsection{The explicit results from recursion formula and comparison with the exact ones}

As for the critical Ising model, in this section we explicitly build the universal function $\mathcal{F}_n (x)$ for various $n$
and at several different orders in the truncations of the Zamolodchikov formula. 
We compare our results with the exact function $\mathcal{F}_n (x)$ for increasing values of $n$.
We also analytically obtain a truncation for the Von Neumann entropy scaling function $F_{\rm vN} (x)$  via replica trick 
and compare it with the simulations in \cite{ATC-CB}.

\begin{figure}[t]
       \centering%
       {\includegraphics[scale=0.55]{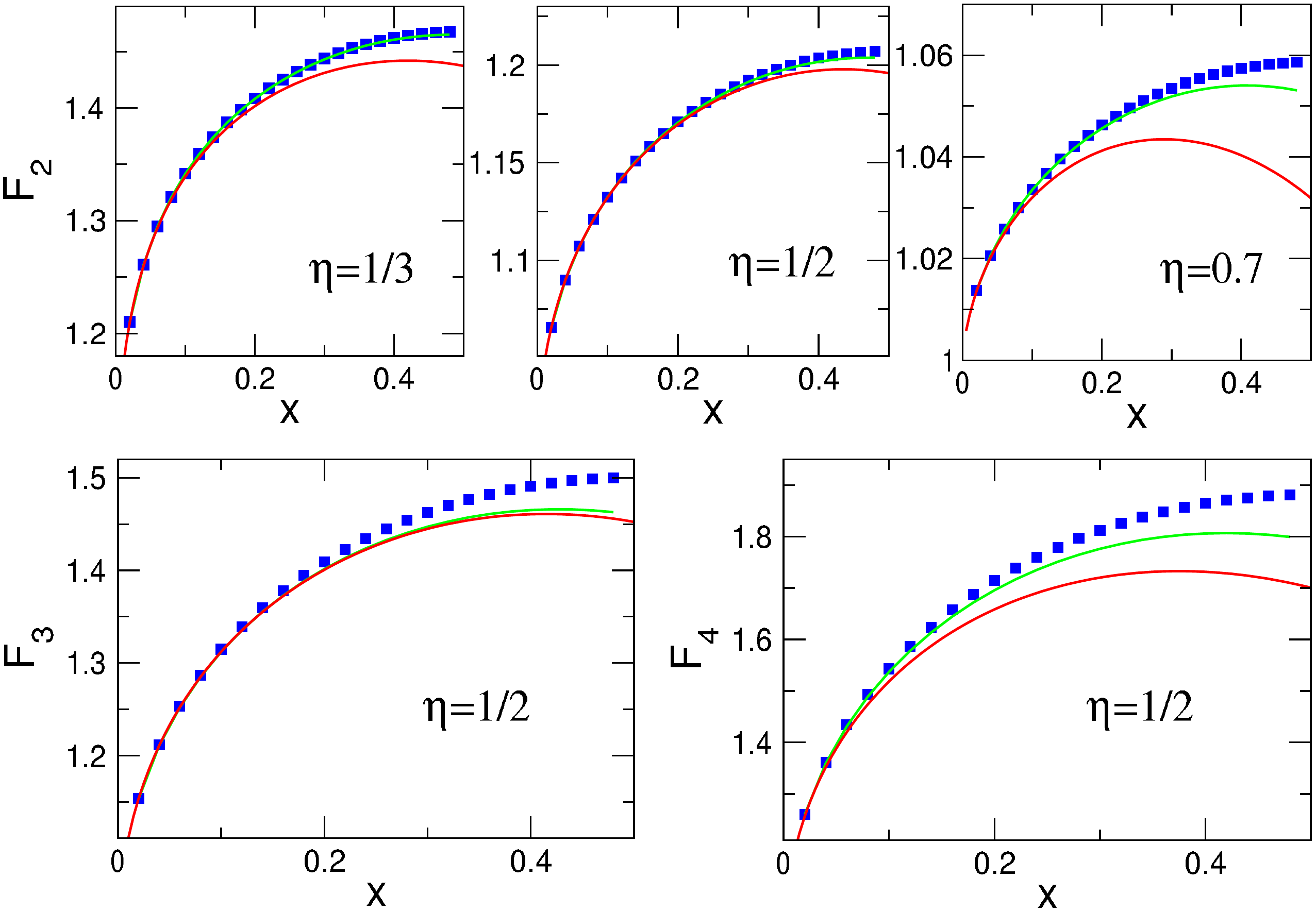}}
        \caption{The function $\mathcal{F}_2 (x)$ for different values of the compactification radius ($\eta= 1/3,  1/2, 0.7$) and the function $\mathcal{F}_3(x)$ and $\mathcal{F}_4(x)$ with $\eta=1/2$ for a compactified boson. In all cases the truncation in the Zamolodchikov formula is at the 0-th order. Two different approximations in the OPE are considered: the fusion channels included are $(0, 0;0, 0) , (1, 0; 1, 0)$ for the red curves and $(0, 0;0, 0) , (1, 0; 1, 0), (1, 0; 2,0) , (0, 1; 0,1), (2, 0; 1, 0), (0, 2; 0, 1)$ for the green curves (with $(p, q; k, l )$ denoting the inclusion of the term in Eq. \eqref{cb-fusionchannel}). The dots represent the exact functions. }
        \label{Fn-Cboson}
  \end{figure}

We first consider the truncation of the Zamolodchikov formula to the first trivial order (i.e. $H \sim 1$ in \cite{GR}) 
and we include the contributions from the first leading conformal blocks. 
The results of this truncation are shown in Figure   \ref{Fn-Cboson}.
In the figure the three panels in the top show ${\cal F}_2(x)$ for three values of $\eta$ while 
the two panels in the bottom display ${\cal F}_3(x)$ and ${\cal F}_4(x)$ at fixed $\eta=1/2$.
The included families for each panel are listed in the caption of the figure.
In all panels the truncated results are compared with the exact results from \cite{CCT-1}. 
It is evident that the approximation improves upon increasing the number of the fusion channels in the OPE (red versus green curves).
It is also to be notice that the quality of the approximation of the function $\mathcal{F}_n$ depends on the value of the parameter $\eta$.
%
Moreover, like for the Ising model,  as $n$ increases a higher number of conformal blocks are required  to well 
approximate $\mathcal{F}_n$.

We also considered the $2$-nd order approximation in the recursion formula.  
However, in this case, the correction to the 0th-order is so small that the two curves are undistinguishable and 
therefore we do not show it here.

\begin{figure}[t]
       \centering
       {\includegraphics[scale=0.4]{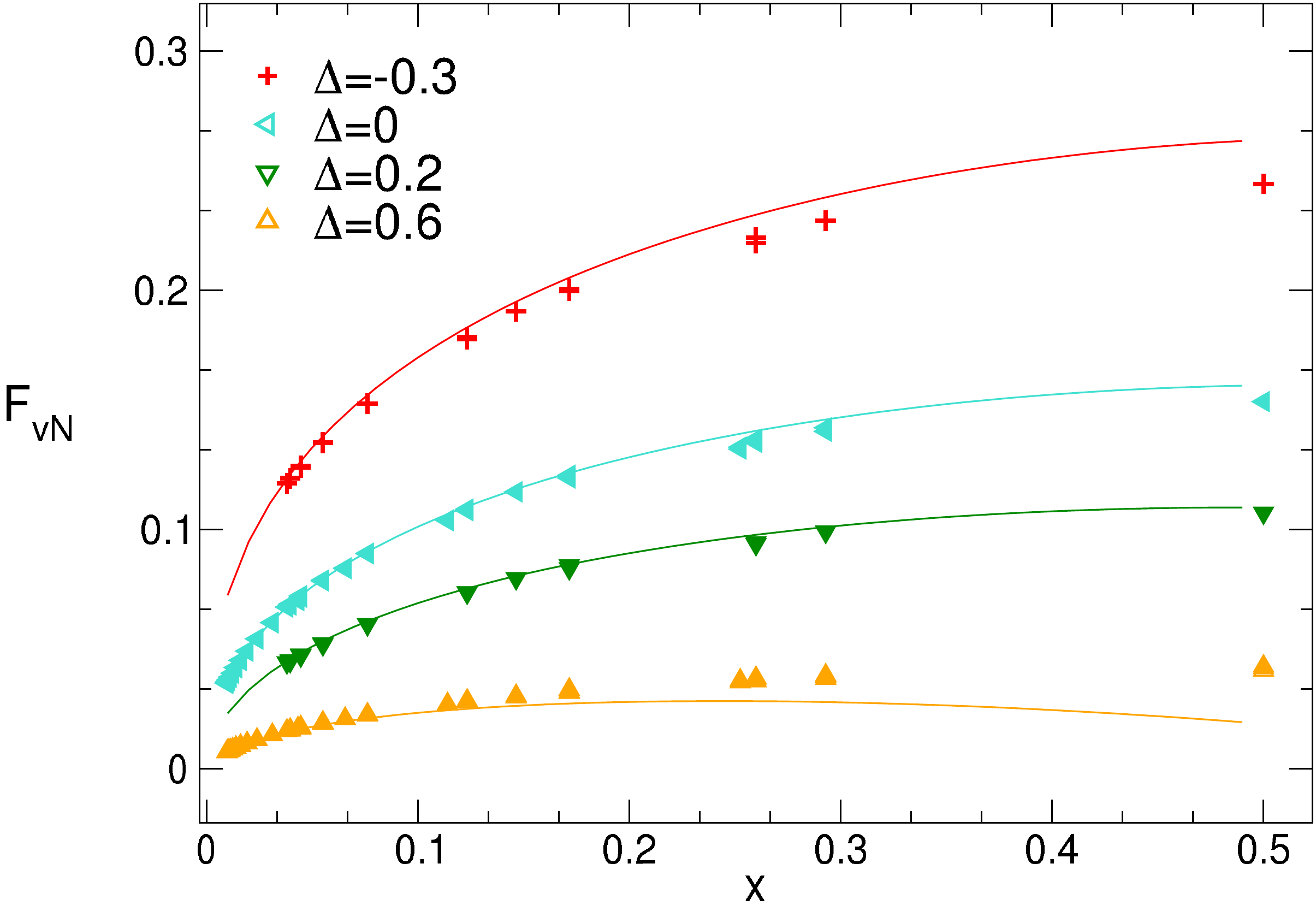}}
        \caption{The continuous lines represent the approximation of the Von Neumann entropy $F_{\rm vN} (x)$ for a compactified boson in Eq. \eqref{FvN-CB-0th}. The dots are the numerics of the XXZ chain in the gapless regime obtained via TTN techniques \cite{ATC-CB}.}
        \label{FvN-CB}
  \end{figure}

Finally, also for the compact boson, we derived the von Neumann entropy via analytic continuation. 
For $\eta <1$, the best approximation we were able to derive is given by
\begin{equation} \label{FvN-CB-0th}
F_{\rm vN}^{(\text{0-th})} (x) = - \frac{5}{3} \log \theta_3 (q) + \frac{1}{12} \log \left( \frac{x(1-x)}{16 q} \right) + s^{(1, 0)'}_{2, 0} (1) (4q)^{\eta} + s^{(1, 0)'}_{4, 0} (1) (4q)^{2 \eta}.
\end{equation}
We report this truncation as function  of $x$ for several values of $\eta$ in Figure \ref{FvN-CB}.
In the Figure, the truncation is compared with numerical simulations presented in Ref. \cite{ATC-CB}. 
These simulations have been obtained from tree tensor networks (TTN) techniques \cite{ATC-CB} of the 
XXZ spin chain with hamiltonian
\begin{equation}
H_{XXZ} (\Delta) = \sum_j [\sigma_j^x \sigma_{j+1}^x + \sigma_j^y \sigma_{j+1}^y + \Delta \, \sigma_j^z \sigma_{j+1}^z   ].
\end{equation}
The model is critical for $\Delta \in [-1, 1]$ and its scaling limit is described by the compact boson
with radius 
\begin{equation}
\eta =2R^2= \frac{\text{Arcos} (- \Delta)}{\pi}.
\end{equation}
Also in this case the agreement for the various $\eta$ is satisfactory. 
Furthermore, the sign of the difference clearly depends on $\eta$.

Surprisingly, for the Von Neumann entropy the introduction of further terms (both in the OPE of twist fields and in the recursion relation) seems to worsen the agreement. The origin of this behaviour is unclear and its understanding  deserves further investigation, in particular in relation to the convergence of the Zamolodchikov series.

\section{Conclusions} \label{section-Conclusions}

In this work we reconsider the approach introduced in Ref. \cite{GR} for the calculation of the entanglement entropy of two disjoint intervals by means of conformal blocks expansion and Zamolodchikov's recursion formula. 
We showed that the inclusion of further fusion channels in the OPE of twist fields in most cases improves the approximation for the scaling functions of the R\'enyi entropies $\mathcal{F}_n (x)$ and of the entanglement entropy $F_{\rm vN} (x)$. 
Moreover, in those cases where the approximation is not good enough, we traced back the origin of the disagreement to the 
truncation of the OPE, rather than to the convergence in the Zamolodchikov's recursion formula for each block 
(which at at the second order appears already very stable). Interestingly, in the only case where the complete form of the OPE is known (i.e., $n=2$ in the Ising model), our approximation perfectly reproduces the exact result \cite{ATC-Ising}.
In this respect a complete classification of the fusion channel appearing in the OPE of twist fields, which is still missing in all the other cases, would be important.

Finally, as a future research direction, it would be interesting to investigate the possibility of using conformal blocks expansion and Zamolodchikov's recursion formula to obtain a feasible truncation of so-called logarithmic negativity \cite{vidal-2002} (related, in the framework of CFT, to the same 4-point correlation function of twist fields, with points ordered in a different way  \cite{CCT-neg2}).
The latter is an entanglement measure in mixed states, that shows an essential singularity for small $x$ 
\cite{CCT-neg2, singularity-negativity}. Such singularity is not yet analytically understood and maybe conformal 
blocks expansion could shed some light on it.

\section*{Aknowledgments} 
We acknowledge very useful discussions with Vincenzo Alba, Davide Bianchini, John Cardy, Olalla Castro-Alvaredo, Benjamin Doyon, Benoit Estienne, Giuseppe Mussardo and in particular Jia-ju Zhang. We also would like to thank Ferdinando Gliozzi and Mohammad Ali Rajabpour for clarifications about their original work \cite{GR}.

\appendix

\section{Vanishing of odd terms in the recursion formula} \label{appendix0}

In this appendix we show that when the Zamolodchikov recursion formula (\ref{Hsum}) is specialised to the case of a
correlation function of four fields with the same conformal dimensions $\Delta_i$ (and therefore same $\lambda_i$), then 
 only the even powers appear, as also noticed in \cite{kt-18}. 
Given the product structure 
\begin{equation} \label{Rmn}
R_{mn}(\tilde{c}, \boldsymbol{\Delta}) = - \frac{1}{2} \prod'_{ab} \frac{1}{\lambda_{ab}} \prod_{p, q} (\lambda_1 + \lambda_2 - \frac{\lambda_{pq}}{2}) (-\lambda_1 + \lambda_2 - \frac{\lambda_{pq}}{2}) (\lambda_3 + \lambda_4 - \frac{\lambda_{pq}}{2}) (\lambda_3 - \lambda_4 - \frac{\lambda_{pq}}{2}) 
\end{equation}
it is sufficient that one term is zero in order for the whole coefficient to vanish.

As first example, let us consider $(m, n) = (1, 1)$, for which the explicit product is 
\begin{equation}
R_{11} ( \tilde{c}  , \boldsymbol{\Delta} ) = - \frac{1}{2} \frac{(\lambda_1 + \lambda_2)( \lambda_2 - \lambda_1 )( \lambda_3  + \lambda_4)( \lambda_3 - \lambda_4) }{\lambda_{01} (\tilde{c}) \; \lambda_{1 0} (\tilde{c}) }.
\end{equation}
Thus if $\lambda_2 = \lambda_1$ or $\lambda_3 = \lambda_4$, then $R_{11} (\tilde{c}, \boldsymbol{\Delta})=0$ and 
therefore $h_1$ vanishes.

What about the other terms? Consider $h_3$ in Eq. \eqref{h3}: the second and third lines vanish because $R_{11}$ is factored out. 
In the first line we should find the coefficients  $R_{13} $ and $R_{31}$. 
For these, among the allowed valued for $p$ and $q$ in the product in Eq. \eqref{Rmn}, there is also the pair $(p, q) = (0, 0)$ 
which makes the whole coefficients vanish, leading  to $h_3=0$. 
Actually, the same reasoning applies to all $h_{2k +1}$: each $R_{mn}$ with odd $(m \cdot n)$ vanishes because of the term 
$(p, q) = (0, 0)$ appearing in the same product. $h_{2k +1}$ is written as sums of terms in which at least one 
$R_{mn}$ with $(m \cdot n)$ odd is factored out. 
%


\section{Approximation method for OPE coefficients} \label{appendix}

In order to get the analytic continuation for the Von Neumann entanglement entropy, the knowledge of the functions $s_{k, l} (n)$ and $s_{k,l}^{(p,q)'}(n)$ (defined for the Ising model and for the compactified boson for $n$ integer, Section \ref{section-Ising} and Section \ref{section-CB} respectively) is not enough. 
In fact, one needs their derivatives with respect to $n$ and therefore their analytic continuation to $n \in \mathbb{R}$.  
However, up to now the analytic continuation is known only for the leading term $s_{2,0} (n)$ in Ising and for 
$s_{(2, 0)}^{(1, 0)}$ in the compactified boson. 
In order to get the contribution of subleading terms, in Ref. \cite{GR} a numerical approximation method 
has been introduced, which we briefly recall below.

The main idea of Ref. \cite{GR} is to approximate the analytic continuations of  
$s_{k, l}(n)$ and $s_{k,l}^{(p,q)'}(n)$ with a polynomial in $n$, of which we know some zeros at integer numbers. 
The unknown coefficients are fixed by fitting the values for the non-zero values in $n$.
This is better understood with a practical example: we write the coefficient $s_{4, 0} (n)$ for the Ising model as
\begin{equation} \label{s-approx}
s_{4, 0} (n) = n (n-1) (n-2) (n-3) (d_0 + d_1 n + d_2 n^2 + d_3 n^3 + \cdots),
\end{equation}
where we used that $s_{4, 0} (n)$  vanishes for $n= 0, 1, 2, 3$.
It is now straightforward  to take the derivative in $n$.

In \cite{GR} the coefficient $s'_{4,0} (1)$ has been found by fitting four free parameters $d_0, d_1, d_2, d_3$.
Here instead, we use the same method but we take into account as many parameters as needed to observe 
convergence for the value of $s'_{k, l} (1)$. 
In practice, we determine $s'_{k, l} (1)$ with varying the number $n_{\rm MAX}$ of terms in the fit.  
We increase $n_{\rm MAX}$ until $s'_{k, l} (1)$ is stable. 
In Table \ref{table-Ising} we report the results (with varying $n_{\rm MAX}$)  for $s'_{4, 0}(1) , s'_{2, 1}(1)$ and $s'_{6, 0}(1)$ for the Ising model, whereas in Table \ref{table-CB} we report the result for $s_{4, 0}^{(1, 0)}{}' (1)$ for the compactified boson, for different values of $\eta$ (aka $\Delta$).

\begin{table}[htpb]
\hspace{-2cm}
\parbox{.6\textwidth}{
\centering
\scalebox{0.8}{
\begin{tabular}{  c | c | c | c  }
    $n_{\rm  MAX}$ &  $s'_{4, 0} (1)$ & $s'_{2, 1} (1)$ & $s'_{6, 0} (1)$  \\ \hline    
    5 & 0.119  & - 0.138  & 0.536  \\
    6 & 0.112 & - 0.134& 0.505   \\
    7 & 0.107 & - 0.131 & 0.486 \\
    8 & 0.104 & - 0.129  & 0.475 \\
    9 & 0.102 & - 0.128 & 0.468 \\
    10 & 0.100 & - 0.127  & 0.471 \\
    11 & 0.099 & - 0.127 & 0.474 \\
    12 & 0.098 & - 0.126 & 0.478  \\
    13 & 0.098 & - 0.126 & 0.482 \\
    14 & 0.098 & - 0.126 & 0.486\\
     15 & 0.098 & - 0.126 & 0.489  \\
      16 & 0.098 & - 0.126 & 0.493 \\
     17 & 0.098 & - 0.126 & 0.496 \\
    \end{tabular} }
        \vspace{0.2cm} 
\caption{Some coefficients $s'_{k, l} (1)$ for the Ising model from Eq. \eqref{s-approx} with $n_{MAX}$ coefficients.}
\label{table-Ising}}
\parbox{.56\textwidth}{
\centering
\scalebox{0.8}{
\begin{tabular}{  c | c | c | c  | c }
    $n_{\rm  MAX}$ & $ \Delta =0$ & $\Delta =-0.3 $ & $   \Delta= 0.2$  & $  \Delta= 0.6$   \\ \hline    
    5 & 0.350  &0. 408 & 0.310 & 0.250  \\
    6 & 0.335 & 0.381 & 0.310 & 0.288   \\
    7 & 0.328 & 0.365 & 0.312 & 0.313 \\
    8 & 0.326 & 0.355  & 0.316 & 0.329  \\
    9 & 0.326 & 0.349 & 0.320 & 0.340 \\
    10 & 0.327 & 0.346  & 0.323 & 0.347 \\
    11 & 0.328 & 0.344 & 0.326 & 0.351 \\
    12 & 0.328 & 0.343 & 0.328 & 0.354  \\
    13 & 0.329 & 0.343 & 0.329 & 0.356 \\
    14 & 0.329 & 0.343 & 0.329 & 0.355 \\
     15 & 0.330 & 0.343 & 0.330 & 0.357  \\
      16 &   &  0.343 & 0.329 & 0.356 \\
     17 &    & 0.343  & 0.329 & 0.356 \\
    \end{tabular} 
  }
\caption{The coefficients $s_{4, 0}^{(1, 0)}{}' (1)$ in the compactified boson for different values of $\Delta$.}
\label{table-CB}}
\end{table}


%
%
%
%
%
%
\end{document}